\documentclass[usenatbib]{mn2e}    

\usepackage{subcaption}
\usepackage[final]{graphicx}
\usepackage{epstopdf}
\usepackage{mathptmx}
\usepackage{float}
\usepackage{fixltx2e}
\usepackage{natbib}
\usepackage{amsmath}
\usepackage{amssymb}
\usepackage[usenames,dvipsnames]{color}
\usepackage{appendix}
\usepackage[mathscr]{euscript}
\usepackage{MnSymbol}

\graphicspath{{figures/}}

\newcommand{\msun}{\mathrm{M}_\odot}
\newcommand{\mstar}{M_\text{star}}
\newcommand{\mhi}{M_\textsc{H\hspace{1pt}i}}

\newcommand{\hi}{\textsc{\mbox{H\hspace{1pt}i}}}
\newcommand{\hii}{\textsc{\mbox{H\hspace{1pt}ii}}}

\newcommand{\eagle}{\textsc{EAGLE}}

\newcommand{\gass}{\textit{GASS}}
\newcommand{\coldgass}{\textit{COLD GASS}}
\newcommand{\bluedisk}{Bluedisk}

\title[\hi{} in \eagle{} galaxies]{The distribution of atomic hydrogen in \eagle{} galaxies: morphologies, profiles, and \hi{} holes}

\author[Y.~Bah\'{e} et al.]{\parbox[t]{\textwidth}{
Yannick~M.~Bah\'{e}\thanks{ybahe@mpa-garching.mpg.de}$^{1}$, Robert A.~Crain$^{2, 3}$, Guinevere Kauffmann$^{1}$, Richard G.~Bower$^{4}$, Joop Schaye$^3$, Michelle Furlong$^{4}$, Claudia Lagos$^{5, 6}$, Matthieu Schaller$^{4}$, James W.~Trayford$^{4}$, Claudio Dalla Vecchia$^{7,8}$, and Tom Theuns$^{4}$}
\vspace*{12pt} \\
$^1$ Max-Planck-Institut f\"{u}r Astrophysik, Karl-Schwarzschild Str. 1, 85748 Garching, Germany\\
$^2$ Astrophysics Research Institute, Liverpool John Moores University, 146 Brownlow Hill, Liverpool, L3 5RF, UK\\
$^3$ Leiden Observatory, Leiden University, PO Box 9513, 2300 RA Leiden, The Netherlands\\
$^4$ Institute for Computational Cosmology, Department of Physics, University of Durham, South Road, Durham DH1 3LE, UK\\
$^5$ European Southern Observatory, Karl-Schwarzschild-Str. 2, 85748 Garching, Germany  \\
$^6$ International Centre for Radio Astronomy Research (ICRAR), M468, University of Western Australia, 35 Stirling Hwy, Crawley, Western Australia, 6009, Australia \\
$^7$ Instituto de Astrof\'{i}sica de Canarias, C/V\'{i}a L\'{a}ctea s/n, E-38205 La Laguna, Tenerife, Spain\\
$^8$ Departamento de Astrof\'{i}sica, Universidad de La Laguna, Av. del Astrof\'{i}sico Francisco S\'{a}nchez s/n, E-38206 La Laguna, Tenerife, Spain\\
}

\begin{document}
\label{firstpage}
\maketitle

\begin{abstract}
We compare the mass and internal distribution of atomic hydrogen (\hi{}) in 2200 present-day central galaxies with $\mstar > 10^{10}\, \msun$ from the 100 Mpc \eagle{} ``Reference'' simulation to observational data. Atomic hydrogen fractions are corrected for self-shielding using a fitting formula from radiative transfer simulations and for the presence of molecular hydrogen using an empirical or a theoretical prescription from the literature. The resulting neutral hydrogen fractions, $M_\text{\hi+H$_2$} / \mstar$, agree with observations to better than 0.1 dex for galaxies with $\mstar$ between 10$^{10}$ and 10$^{11}$ $\msun$. Our fiducial, empirical H$_2$ model based on gas pressure results in galactic \hi{} mass fractions, $\mhi / \mstar$, that agree with observations from the \gass{} survey to better than 0.3 dex, but the alternative theoretical H$_2$ formula from high-resolution simulations leads to a negative offset in $\mhi / \mstar$ of up to 0.5 dex. Visual inspection of mock \hi{} images reveals that most \hi{} disks in simulated \hi-rich galaxies are vertically disturbed, plausibly due to recent accretion events. Many galaxies (up to 80 per cent) contain spuriously large \hi{} holes, which are likely formed as a consequence of the feedback implementation in \eagle{}. The \hi{} mass--size relation of all simulated galaxies is close to (but 16 per cent steeper than) observed, and when only galaxies without large holes in the \hi{} disc are considered, the agreement becomes excellent (better than 0.1 dex). The presence of large \hi{} holes also makes the radial \hi{} surface density profiles somewhat too low in the centre, at $\Sigma_\hi > 1 \msun\, \text{pc}^{-2}$ (by a factor of $\lesssim 2$ compared to data from the \bluedisk{} survey). In the outer region ($\Sigma_\hi < 1 \msun\, \text{pc}^{-2}$), the simulated profiles agree quantitatively with observations. Scaled by \hi{} size, the simulated profiles of \hi-rich ($\mhi > 10^{9.8} \msun$) and control galaxies ($10^{9.1}\, \msun > \mhi > 10^{9.8}\, \msun$) follow each other closely, as observed. 
\end{abstract}



\section{Introduction}
\label{sec:introduction}

Radio observations have revealed the presence of atomic hydrogen (\hi) in the Milky Way (see \citealt{Dickey_Lockman_1990}), as well as in many other galaxies (see \citealt{Walter_et_al_2008} and references therein). Although the median \hi{} mass fraction $\mhi/\mstar$ is only $\sim$0.1 for Milky Way mass galaxies ($\mstar \approx 10^{10.5} \msun$), the presence of substantial scatter means that the ratio can exceed unity in individual cases \citep{Catinella_et_al_2010}. This \hi{} reservoir is believed to be fuel for future star formation (e.g.~\citealt{Prochaska_Wolfe_2009, Dave_et_al_2010, van_de_Voort_et_al_2012}), which makes the ability to correctly model its structure and evolution an integral part of the wider quest to better understand galaxy formation.

The long timescales of galaxy formation ($\gtrsim 1$ Gyr) imply that observations are effectively limited to one point in time for any individual galaxy, so that studying the evolution of galactic gas necessarily involves theoretical modelling. In the ``Semi-Analytic Modelling'' (SAM) approach (e.g.~\citealt{Kauffmann_et_al_1993, Guo_et_al_2011,Fu_et_al_2013}), the evolution of baryonic galaxy components is described by analytic equations that are combined with an  underlying dark matter distribution from N-body simulations (e.g.~\citealt{Springel_et_al_2005, Boylan-Kolchin_et_al_2009}) or the extended Press-Schechter formalism \citep{Bond_et_al_1991, Bower_1991}. SAMs have become increasingly refined over time, and a number of authors have used them to study various aspects of \hi{} in galaxies such as its evolution \citep{Lagos_et_al_2011, Popping_et_al_2014}, radial distribution \citep{Fu_et_al_2013, Wang_et_al_2014} and origin in early-type galaxies \citep{Lagos_et_al_2014}. 

However, SAMs are not able to predict the detailed structure of gas within and around galaxies: its accretion, for example, is typically modelled in an ad-hoc way without fully accounting for the filamentary structure of the intergalactic medium (but see \citealt{Benson_Bower_2010} for a counter-example). This motivates the use of cosmological hydrodynamical simulations, which model the accretion and outflows of gas from galaxies self-consistently, and at (potentially) high spatial resolution. Further benefits include the ability to trace the thermodynamic history of individual fluid elements (e.g.~\citealt{Keres_et_al_2005, van_de_Voort_et_al_2011, Nelson_et_al_2013}), and that they permit the study of satellite galaxies without additional assumptions (e.g.~\citealt{Bahe_McCarthy_2015}).

A number of authors have studied low-redshift \hi{} with cosmological hydrodynamical simulations in the past. \citet{Popping_et_al_2009} successfully reproduced the observed distribution of \hi{} column densities over seven orders of magnitude, as well as the \hi{} two-point correlation function (see also \citealt{Duffy_et_al_2012, Rahmati_et_al_2013}). The sensitivity of \hi{} to supernova feedback, and its evolution over cosmic time, was explored by \citet{Dave_et_al_2013}, while \citet{Cunnama_et_al_2014} and \citet{Rafieferantsoa_et_al_2015} investigated the influence of the group/cluster environment on \hi{}. 

However, a common problem of these simulations has been their inability to produce galaxies whose stellar component agrees with observations. In particular, angular momentum from infalling gas was typically dissipated too quickly and too severely to form realistic discs \citep{Steinmetz_Navarro_1999}, and `overcooling' (e.g.~\citealt{Katz_et_al_1996}) manifested itself in galaxy stellar mass functions that are too high at the massive end (e.g.~\citealt{Crain_et_al_2009, Oppenheimer_et_al_2010, Lackner_et_al_2012}). 

In the recent past, several groups have developed simulations which are able to avoid these problems. Incorporation of efficient supernova feedback --- in a physical and numerical sense --- and/or increased resolution has led to the formation of realistic disk galaxies (e.g.~\citealt{Governato_et_al_2007, Governato_et_al_2010, McCarthy_et_al_2012b, Aumer_et_al_2013, Marinacci_et_al_2014}). The inclusion of additional feedback from accreting super-massive black holes (`AGN feedback'), on the other hand, has reduced the overcooling problem at the high-mass end and led to more accurate stellar masses of simulated galaxies (\citealt{Rosas-Guevara_et_al_2013}; see also \citealt{Springel_DiMatteo_Hernquist_2005, Sijacki_et_al_2007, Booth_Schaye_2009, Vogelsberger_et_al_2013}). With these and other improvements, the \eagle{} project \citep{Schaye_et_al_2015, Crain_et_al_2015} has yielded a cosmologically representative population of galaxies with realistic properties such as stellar masses and sizes (see also \citealt{Vogelsberger_et_al_2014} for the \textsc{Illustris} simulation). \eagle{} has also been shown to broadly reproduce e.g.~the observed colour distribution of galaxies \citep{Trayford_et_al_2015} at $z \sim 0$, as well as the redshift evolution of the stellar mass growth and star formation rates \citep{Furlong_et_al_2015}.

The Lagrangian Smoothed Particle Hydrodynamics (SPH) formalism adopted by \eagle{} makes it, in principle, possible to study directly the physics governing the accretion and outflow of atomic hydrogen in simulated galaxies. Unlike the $z \approx 0$ stellar mass function and sizes, \hi{} properties were not taken into account when calibrating the \eagle{} galaxy formation model. It is therefore uncertain whether the distribution of \hi{} is modelled correctly: as \citet{Crain_et_al_2015} have shown, even stellar masses and sizes are sufficiently independent of each other that reproducing observations of one does not necessarily imply success with the other. Comparing the \hi{} properties of simulated galaxies to observations therefore offers an opportunity to directly test the galaxy formation model, as well as being a necessary step to ascertain the extent to which simulation predictions are trustworthy. 

At high redshift ($z \geq 1$), \citet{Rahmati_et_al_2015} have shown that the column density distribution function and covering fractions of \hi{} absorbers in \eagle{} agree with observations, while \citet{Lagos_et_al_2015} demonstrated that \eagle{} galaxies contain realistic amounts of molecular hydrogen (H$_2$) both at $z=0$ and across cosmic history. Here, we conduct a series of detailed like-with-like comparisons between \eagle{} and recent low-redshift \hi{} observations including the \emph{Galex Arecibo SDSS Survey} (\gass; \citealt{Catinella_et_al_2010, Catinella_et_al_2013}) and the Bluedisk project \citep{Wang_et_al_2013, Wang_et_al_2014}. Our aim is to analyse the distribution of \hi{} within individual $z=0$ galaxies; the cosmological distribution of \hi{} in \eagle{} will be investigated separately (Crain et al., in prep.).

The remainder of this paper is structured as follows. In Section \ref{sec:eagle}, we review the key characteristics of the \eagle{} project, and give an overview of the \gass{} \citep{Catinella_et_al_2010} and \coldgass{} \citep{Saintonge_et_al_2011} surveys in Section \ref{sec:surveys}. Our \hi{} modelling scheme is then described in Section \ref{sec:himodelling}, followed by a comparison of galaxy-integrated neutral hydrogen and \hi{} masses to observations in Section \ref{sec:gass}. Section \ref{sec:profiles} analyses the internal distribution of \hi{} in the simulated galaxies, including a comparison of \hi{} surface density profiles to \bluedisk{} data. Our results are summarised and discussed in Section \ref{sec:summary}. All masses and distances are given in physical units unless specified otherwise. A flat $\Lambda$CDM cosmology with Hubble parameter $h \equiv $ H$_{0}/(100\,{\rm km}\,{\rm s}^{-1}{\rm Mpc}^{-1}) = 0.6777$, dark energy density parameter $\Omega_\Lambda = 0.693$ (dark energy equation of state parameter $w=-1$), and matter density parameter $\Omega_{\rm M} = 0.307$ as in \citet{Planck_2014} is used throughout this paper. The \eagle{} simulations adopt a universal \citet{Chabrier_2003} stellar initial mass function (IMF) with minimum and maximum stellar masses of 0.1 and 100 $\msun$, respectively.


\section{The \eagle{} simulations}
\label{sec:eagle}

\subsection{Simulation characteristics}
The ``Evolution and Assembly of GaLaxies and their Environments'' (\eagle) project consists of a large suite of many cosmological hydrodynamical simulations of varying size, resolution and sub-grid physics prescriptions. They are introduced and described in detail by \citet{Schaye_et_al_2015} and \citet{Crain_et_al_2015}; here we only summarise the main characteristics that are particularly relevant to our study. 

The largest simulation (Ref-L100N1504 in the terminology of \citealt{Schaye_et_al_2015}), upon which our analysis here is based, fills a cubic box of side length 100 comoving Mpc (`cMpc') with $N = 1504^3$ dark matter particles ($M_\text{DM} = 9.7 \times 10^6\, \msun$) and an initially equal number of gas particles ($M_\text{gas} = 1.81 \times 10^6\, \msun$). The simulation was started at $z = 127$ from cosmological initial conditions \citep{Jenkins_2013}, and evolved to $z = 0$ using a modified version of the \textsc{gadget-3} code \citep{Springel_2005}. These modifications include a number of hydrodynamics updates collectively referred to as ``Anarchy'' (Dalla Vecchia, in prep.; see also \citealt{Hopkins_2013},  Appendix A of \citealt{Schaye_et_al_2015}, and \citealt{Schaller_et_al_2015}) which eliminate most of the problems associated with ``traditional'' SPH codes related to the treatment of surface discontinuities (e.g.~\citealt{Agertz_et_al_2007, Mitchell_et_al_2009}) and artificial gas clumping (e.g.~\citealt{Nelson_et_al_2013}).

The gravitational softening length is 0.7 proper kpc (`pkpc') at redshifts $z < 2.8$, and 2.66 ckpc at earlier times. In the warm interstellar medium, the Jeans scales are therefore marginally resolved, but the same is not true for the cold molecular phase. For this reason, the simulation imposes a temperature floor $T_\text{eos} (\rho)$  on gas with $n_\text{H} > 0.1$ cm$^{-3}$, in the form of a polytropic equation of state $P \propto \rho^\gamma$ with index $\gamma = 4/3$ and normalised to $T_\text{eos} = 8 \times 10^3$ K at $n_\text{H} = 10^{-1} \text{cm}^{-3}$ (see \citealt{Schaye_DallaVecchia_2008} and  \citealt{DallaVecchia_Schaye_2012} for further details). In addition, gas at densities $n_\text{H} \geq 10^{-5}$ cm$^{-3}$ is prevented from cooling below 8000 K. 

The \eagle{} simulation code includes significantly improved sub-grid physics prescriptions. These include element-by-element radiative gas cooling \citep{Wiersma_et_al_2009a} in the presence of the Cosmic Microwave Background (CMB) and an evolving \citet{Haardt_Madau_2001} UV/X-ray background, reionization of hydrogen at $z = 11.5$ and helium at $z \approx 3.5$ \citep{Wiersma_et_al_2009b}, star formation implemented as a pressure law \citep{Schaye_DallaVecchia_2008} with a metallicity-dependent density threshold \citep{Schaye_2004}, stellar mass loss and chemical enrichment on an element-by-element basis \citep{Wiersma_et_al_2009b}, as well as energy injection from supernovae \citep{DallaVecchia_Schaye_2012} and accreting supermassive black holes (AGN feedback; \citealt{Rosas-Guevara_et_al_2013, Schaye_et_al_2015}) in thermal form. 

For a detailed description of how these sub-grid models are implemented in \eagle{}, the interested reader is referred to \citet{Schaye_et_al_2015}. However, three aspects in the implementation of energy feedback from star formation merit explicit mention here. Firstly, because the feedback efficiency cannot be predicted from first principles, its strength was calibrated to reproduce the $z\approx 0$ galaxy stellar mass function and sizes. Secondly, the feedback parameterisation depends only on local gas quantities, in contrast to e.g.~the widely-used practice of scaling the parameters with the (global) velocity dispersion of a galaxy's dark matter halo (e.g.~\citealt{Okamoto_et_al_2005, Oppenheimer_Dave_2006, Dave_et_al_2013, Vogelsberger_et_al_2013,  Puchwein_Springel_2013}). Finally, star formation feedback in \eagle{} is made efficient not by temporarily disabling hydrodynamic forces or cooling for affected particles (e.g.~\citealt{Springel_Hernquist_2003, Stinson_et_al_2006}), but instead by stochastically heating a small number of particles by a temperature $\Delta T = 10^{7.5}\, \text{K}$ \citep{DallaVecchia_Schaye_2012}. These details can be expected to influence in non-trivial ways the galactic distribution of \hi{} (see e.g.~\citealt{Dave_et_al_2013}), so that an examination of this diagnostic also informs our understanding of the impact of this scheme on the structure of the simulated ISM.

\subsection{Galaxy selection}
From the 100 cMpc \eagle{} simulation Ref-L100N1504, we select our $z = 0$ target galaxies as self-bound subhaloes --- identified using the \textsc{subfind} algorithm (\citealt{Dolag_et_al_2009}; see also \citealt{Springel_et_al_2001b}) --- with a stellar mass of $\mstar \geq 10^{10}\, \msun$. This limit ensures that individual galaxies are well resolved ($\gg$ 1000 baryon particles) and that our sample is directly comparable to the observational \gass{} and Bluedisk surveys. Crain et al. (in prep.) will present the full \hi{} mass function in \eagle{} extending down to much smaller galaxies. Stellar masses are computed as the total mass of all gravitationally bound star particles within a spherical aperture of 30 kpc, centered on the particle for which the gravitational potential is minimum. \citet{Schaye_et_al_2015} showed that this definition mimics the Petrosian mass often used by optical surveys.   

Note that we only select central galaxies --- i.e.~the most massive subhalo in a friends-of-friends halo --- because satellites are subject to additional complex environmental processes that can impact upon their \hi{} content (e.g.~\citealt{Fabello_et_al_2012, Catinella_et_al_2013, Zhang_et_al_2013}; see also \citealt{Bahe_et_al_2013}). We focus here on testing the arguably more fundamental accuracy of the simulations for centrals; the \hi{} properties of \eagle{} satellites will be discussed elsewhere (Marasco et al., in prep.). In total, we have a sample of 2200 galaxies, the vast majority of which (2039) have stellar masses below $10^{11} \, \msun$. 


\section{The \gass{} and \coldgass{} surveys}
\label{sec:surveys.gass}
\label{sec:surveys}
Before describing our \hi{} analysis as applied to \eagle{}, we now give a brief overview of the \gass{} and \coldgass{} surveys, which will be compared to our simulations below. We also describe our approach for comparing \eagle{} in a consistent way to these observations. For clarity, we will describe the third main survey used in our work, \bluedisk{} \citep{Wang_et_al_2013}, in Section \ref{sec:profiles.comparison} where its results are compared to predictions from \eagle.

\subsection{The Galex-Arecibo-SDSS Survey (\gass)}
The Galex-Arecibo-SDSS Survey\footnote{Data available at http://www.mpa-garching.mpg.de/GASS} (\gass; \citealt{Catinella_et_al_2010, Catinella_et_al_2013}) was designed to provide an unbiased census of the total \hi{} content in galaxies with stellar mass $\mstar > 10^{10} \msun$. Measuring this observationally does not require very high spatial resolution and can therefore be achieved with single-dish observations on e.g.~the Arecibo telescope. However, an important issue is that of galaxy selection: blind surveys such as \mbox{ALFALFA} \citep{Giovanelli_et_al_2005} are naturally more likely to detect abnormally \hi-rich than -poor galaxies because the volume over which the latter can be detected is small \citep{Catinella_et_al_2010, Huang_et_al_2012}. The galaxies in \gass{} are therefore selected only by stellar mass, and observed until either the 21-cm line from \hi{} is detected or an upper limit of $M_\hi / \mstar \approx 0.015$ has been reached\footnote{This limit is fixed to $\mhi = 10^{8.7}\, \msun$ for galaxies with $\mstar < 10^{10.5}\, \msun$, so that the gas fraction detection threshold increases towards lower stellar masses.}. Out of the 760 galaxies in the full \gass{} sample, centrals are selected by cross-matching to the \citet{Yang_et_al_2012} SDSS group catalogue (see \citealt{Catinella_et_al_2013} for details), which leaves us with 386 galaxies with $10^{10}\,\, \msun \leq \mstar \leq 10^{11} \msun$ (and 522 at $\mstar \geq 10^{10} \msun$). We note that the equivalent \eagle{} sample is almost an order of magnitude larger ($N = 2083$ and 2200, respectively), because of the larger effective volume. 

In order to make the comparison between \eagle{} and \gass{} fair, it is important to compute $\mhi$ for \eagle{} galaxies as done in observations, i.e.~by integrating over the same range in projected radius and line-of-sight distance. For the former, we use a fixed value of 70 kpc, which roughly corresponds to the \emph{Arecibo L-Band Feed Array} (ALFA) FWHM beam size of $\sim$3.5 arcmin \citep{Giovanelli_et_al_2005} at the median redshift of the \gass{} sample, $\tilde{z} = 0.037$ \citep{Catinella_et_al_2010}. The line of sight is taken as the simulation $z$-coordinate; we include all particles (including those outside haloes) with peculiar velocity relative to the mass-weighted velocity of the galaxy subhalo in the range [-400, +400] km s$^{-1}$ to approximately match what was done in \gass{}. A comparison of this integration range with simple spherical shell apertures can be found in Appendix \ref{sec:app.aperture}, which confirms that masses obtained with this ``\gass-equivalent'' method agree well with the mass of \hi{} inside a (3D) aperture of 70 kpc, but exceed those measured inside a 30 kpc aperture at both the high- and low-$\mhi$ end by typically up to a factor of two.

\subsection{CO Legacy Database for GASS (\coldgass)}
To complement the \gass{} database with information on the molecular hydrogen content of galaxies, the \coldgass{}\footnote{Data available at http://www.mpa-garching.mpg.de/COLD\texttt{\char`_}GASS} survey \citep{Saintonge_et_al_2011} observed a randomly selected subset of $\sim$250 galaxies from the \gass{} sample in CO with the IRAM 30-m telescope. Similarly to \gass{}, galaxies were observed until either the CO (1--0) line was detected or an upper limit equivalent to an H$_2$ mass fraction of $\sim$1.5 per cent (for galaxies with $\mstar > 10^{10.6} \msun$) or an absolute H$_2$ mass of  $10^{8.8} \msun$ (for galaxies with $\mstar < 10^{10.6} \msun$) was achieved.

A detailed comparison of \eagle{} to results from \coldgass{} is presented by \citet{Lagos_et_al_2015}. Here, we combine the results from \gass{} and \coldgass{} to obtain observational constraints on the \emph{total} neutral hydrogen mass in galaxies, which we compare to predictions from \eagle{} in Section \ref{sec:neutralhydrogencomp}. For simplicity, we adopt the same particle selection as described above for \gass{}: this is justified because H$_2$ is concentrated more strongly towards the galaxy centre than \hi{} and the \coldgass{} survey is designed to measure the total H$_2$ masses of its galaxies (see \citealt{Saintonge_et_al_2011} for more details). Neutral hydrogen masses obtained from the simulations with the relatively large aperture matched to \gass{} can therefore be meaningfully compared to the sum of \hi{} and H$_2$ masses from \gass{} and \coldgass, respectively.


\section{\hi{} modelling}
\label{sec:himodelling}
The \eagle{} simulation output itself only contains the mass of hydrogen in each gas particle, but not how much of this is in ionised (\hii), atomic (\hi) or molecular (H$_2$) form\footnote{The radiative cooling prescription of \eagle{} takes into account that only a fraction of the gas is neutral. However, these ratios are computed without accounting for self-shielding (see \citealt{Rahmati_et_al_2013} and \citealt{Schaye_et_al_2015} for further details) and can therefore not be used directly in our present work, where we study the highly self-shielded regime of galaxy interiors.}. Although it is possible to separate these self-consistently using radiation transport and detailed chemical network modelling (e.g.~\citealt{Pawlik_et_al_2008, Altay_et_al_2011,Christensen_et_al_2012, Rahmati_et_al_2013, Richings_et_al_2014a, Richings_et_al_2014b, Walch_et_al_2015}), the computational expense of dynamically coupling these techniques to the simulation and e.g.~calculate star formation rates directly from the H$_2$ phase is unfeasibly high for a 100 cMpc simulation like \eagle{}. Although we cannot, therefore, make truly self-consistent predictions for the individual hydrogen phases, we can still gain insight by employing an approximation scheme in post-processing to calculate the \hi{} mass of gas particles, as follows.

\subsection{Neutral and ionised hydrogen}
\label{sec:neutralh}
First, we compute the fraction of hydrogen in each gas particle that is \emph{neutral} (\hi{} and H$_2$). For this, we use the ionisation fitting formula of \citet{Rahmati_et_al_2013}, which was calibrated using (smaller) simulations with detailed radiation transport modelling\footnote{For particles within 0.5 dex of the imposed equation of state, we assume a fixed temperature of $T = 10^4$ K when calculating collisional ionisation and recombination with this prescription.}. Their prescription relates the total ionisation rate (photo- plus collisional ionisation) to that from the UV background, for which we adopt a value\footnote{This value is larger by a factor of $\sim$3 than the more recent determination by \citet{Haardt_Madau_2012}. We have tested both, and found no significant impact on the \hi{} results presented here.} of $\Gamma_\text{UVB} = 8.34 \times 10^{-14}\,\, \text{s}^{-1}$ \citep{Haardt_Madau_2001}, accounting for self-shielding. Not taken into account, however, is the (difficult to constrain) effect of local stellar radiation, which \citet{Rahmati_et_al_2013b} found to affect dense \hi{} systems even at $z=0$. In Section \ref{sec:neutralhydrogencomp}, we show that the resulting neutral gas masses are in good agreement with observational constraints.

\subsection{Atomic and molecular hydrogen (\hi{}/\textsc{H}$_2$)}
In a second step, we then model the fractions of neutral hydrogen in molecular (H$_2$) and atomic form (\hi) with two different approaches. Our fiducial method, similar to what was done by \citet{Altay_et_al_2011}, \citet{Duffy_et_al_2012}, and \citet{Dave_et_al_2013}, is to exploit the empirical relation between gas pressure and molecular fraction \citep{Wong_Blitz_2002, Blitz_Rosolowski_2006} which can be measured observationally on scales comparable to the resolution of \eagle. This approach is approximately self-consistent, because star formation is also implemented based on pressure \citep{Schaye_DallaVecchia_2008} and observational evidence strongly suggests a link between the two (e.g.~\citealt{Leroy_et_al_2008, Krumholz_et_al_2009, Bigiel_et_al_2011, Huang_Kauffmann_2014}; but see the theoretical work of \citealt{Glover_Clark_2012}).

From observations of 11 nearby, non-interacting galaxies spanning almost one decade in total metallicity and three decades in pressure, \citet[BR06]{Blitz_Rosolowski_2006} derived the H$_2$/\hi{} fraction in terms of the mid-plane gas pressure $P$ as 
\begin{equation}
R_\text{mol} \equiv \frac{\Sigma_{\text{H}_2}}{\Sigma_\hi} = \left(\frac{P}{P_0}\right)^{\alpha},
\label{eq:pressurelaw}
\end{equation} 
with best fit parameters\footnote{\citet{Leroy_et_al_2008} studied a somewhat larger sample of 23 galaxies, and found a best-fit normalisation $P_0 / k_B = 1.7 \times 10^4$ cm$^{-3}$ K and exponent $\alpha = 0.8$; this parameterisation was used by \citet{Duffy_et_al_2012} and \citet{Dave_et_al_2013}. We show in Appendix \ref{sec:app.h2} that the difference between applying these two parameterisations to our simulated galaxies is negligible. The same is true for the effect of including three additional interacting galaxies in the analysis of BR06.} $P_0/k_B = 4.3 \times 10^4\,\,  \text{cm}^{-3}\,\, \text{K}$ and $\alpha = 0.92$. Assuming that the molecular and atomic phases have the same scale-height, $R_\text{mol}$ is also equal to the ratio between the \emph{volume} densities of H$_2$ and \hi. We furthermore assume that neutral hydrogen only has a contribution from H$_2$ in particles with a non-zero star formation rate, the density threshold for which is motivated by whether physical conditions allow the formation of a cold molecular phase \citep{Schaye_2004}. 

As an alternative, we also consider the theoretically motivated H$_2$ partitioning scheme of \citet[GK11]{Gnedin_Kravtsov_2011}, which is based on high-resolution simulations with an explicit treatment of the formation and destruction of H$_2$. For more details on this scheme and its implementation in \eagle, we refer the interested reader to \citet{Lagos_et_al_2015}, where this prescription was shown to yield good agreement between the H$_2$ content of \eagle{} galaxies and observations. Further approaches of modelling the \hi{} in \eagle{} are explored in Appendix \ref{sec:app.h2}; these include simple prescriptions such as ignoring H$_2$ altogether or assuming a fixed ratio of $m_{\text{H}_2}/m_\hi = 0.3$ for each particle (as in \citealt{Popping_et_al_2009}), both of which give similar results as our fiducial empirical BR06 method described above. In the same place, we also test the alternative theoretical prescription by \citet{Krumholz_2013} as implemented into \eagle{} by \citet{Lagos_et_al_2015}.


\section{Neutral and atomic hydrogen fractions compared to observations}
\label{sec:gass}

\subsection{Neutral hydrogen fractions}
\label{sec:neutralhydrogencomp}
\begin{figure}
  \centering
    \includegraphics[width=\columnwidth]{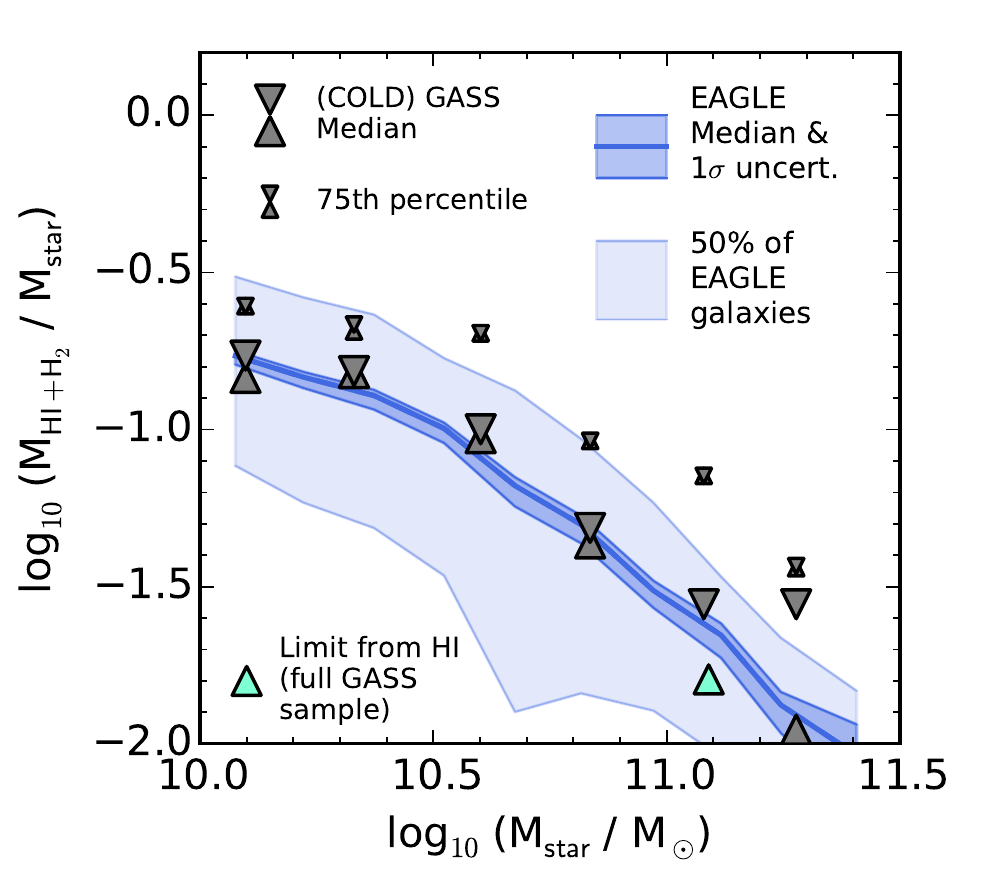}
       \caption{Neutral hydrogen mass fractions for our simulated galaxies as predicted by the \citet{Rahmati_et_al_2013} fitting formula (blue line, shaded bands show the 1$\sigma$ uncertainty and 50 per cent scatter, respectively). For comparison, observational data from the combined \gass{} and \coldgass{} surveys are shown as grey symbols, upward (downward) facing triangles differ in that non-detections are set to zero (upper limits). Large triangles show the observed medians, small ones the 75th percentile of the distribution. The light blue triangle shows an additional lower limit from \hi{} masses in the full \gass{} survey (see text for details). The neutral hydrogen masses in \eagle{} agree with observational constraints to within 0.1 dex, although there are large uncertainties on the observational median at $\mstar > 10^{11} \msun$.}
       \label{fig:neutralgas}
  \end{figure}

We begin by showing in Fig.~\ref{fig:neutralgas} the total neutral hydrogen fraction, i.e.~$M_\text{\hi + H$_2$} / \mstar$, of our simulated galaxies as a function of stellar mass $\mstar$ (blue). The solid dark blue line shows the running median of the distribution. The dark and light shaded bands indicate the statistical 1$\sigma$ uncertainty on the median and the 50 per cent scatter, respectively, i.e.~they extend from $f_\text{low}$ to $f_\text{high}$ where $f = M_\text{\hi+H$_2$}/\mstar$ and $f_\text{low (high)} = \tilde{f} + (P_\text{15.9 (84.1)}-\tilde{f}) / \sqrt{N}$; $\tilde{f}$ here denotes the median and $P_n$ the $n$th percentile of the distribution in a bin with $N$ galaxies. This prediction is compared to observational constraints from the intersection of the \gass{} and \coldgass{} surveys shown as grey symbols. Both have a large fraction of non-detections: only 46 per cent of central galaxies targeted in both surveys are detected in \hi{} and CO, although the majority of galaxies (83 per cent) are detected in at least one component. To bracket the resulting uncertainty on the observed median, we have computed it with non-detections set both to zero (giving lower limits, shown by upward facing triangles) and the observational upper limit (downward facing triangles). At $\mstar < 10^{11} \msun$, both approaches differ by less than 0.2 dex. The 75th percentile of the observed distribution is analogously shown by small triangles. The impact of non-detections is much smaller here ($< 0.1$ dex).

The median neutral hydrogen fraction predicted by \eagle{} agrees remarkably well with observational constraints, deviating by $\lesssim 0.1$ dex in the regime $\log_{10} (\mstar/\msun) = [10.0, 11.0]$ in the sense that the simulated galaxies contain, in general, slightly too little neutral gas. The 75th percentiles agree at a similar level, but without a consistent sign of the deviation. For the most massive galaxies ($\mstar > 10^{11} \msun$), the large observational uncertainties induced by frequent non-detections prevent strong statements on the accuracy of the simulation prediction for the median neutral fraction, but the 75th percentiles are well-constrained observationally and show a significant shortfall of the simulation, by $\sim$0.3 dex.

We note that in the second most massive $\mstar$ bin, $\log_{10} (\mstar/\msun) = [11.0, 11.25]$, less than 50 per cent of (central) galaxies in the \coldgass{} sample are detected in either \hi{} or CO, so the log-scaling of Fig.~\ref{fig:neutralgas} prevents us from showing a lower limit on the observed median here. However, in the (larger) \gass{} sample, the \hi{} detection fraction in the same bin is 52 per cent, so we can place at least a (conservative) lower limit on the neutral gas fraction in this bin from the \gass{} \hi{} median alone (light blue triangle in Fig.~\ref{fig:neutralgas}). Including this additional constraint, the median \eagle{} neutral gas fractions are consistent with observations at the 0.2 dex level over the range $\log_{10} (\mstar/\msun) = [10.0, 11.25]$.

\subsection{Atomic hydrogen fractions compared to \gass{}}
\begin{figure*}
  \centering
    \includegraphics[width=2.13\columnwidth]{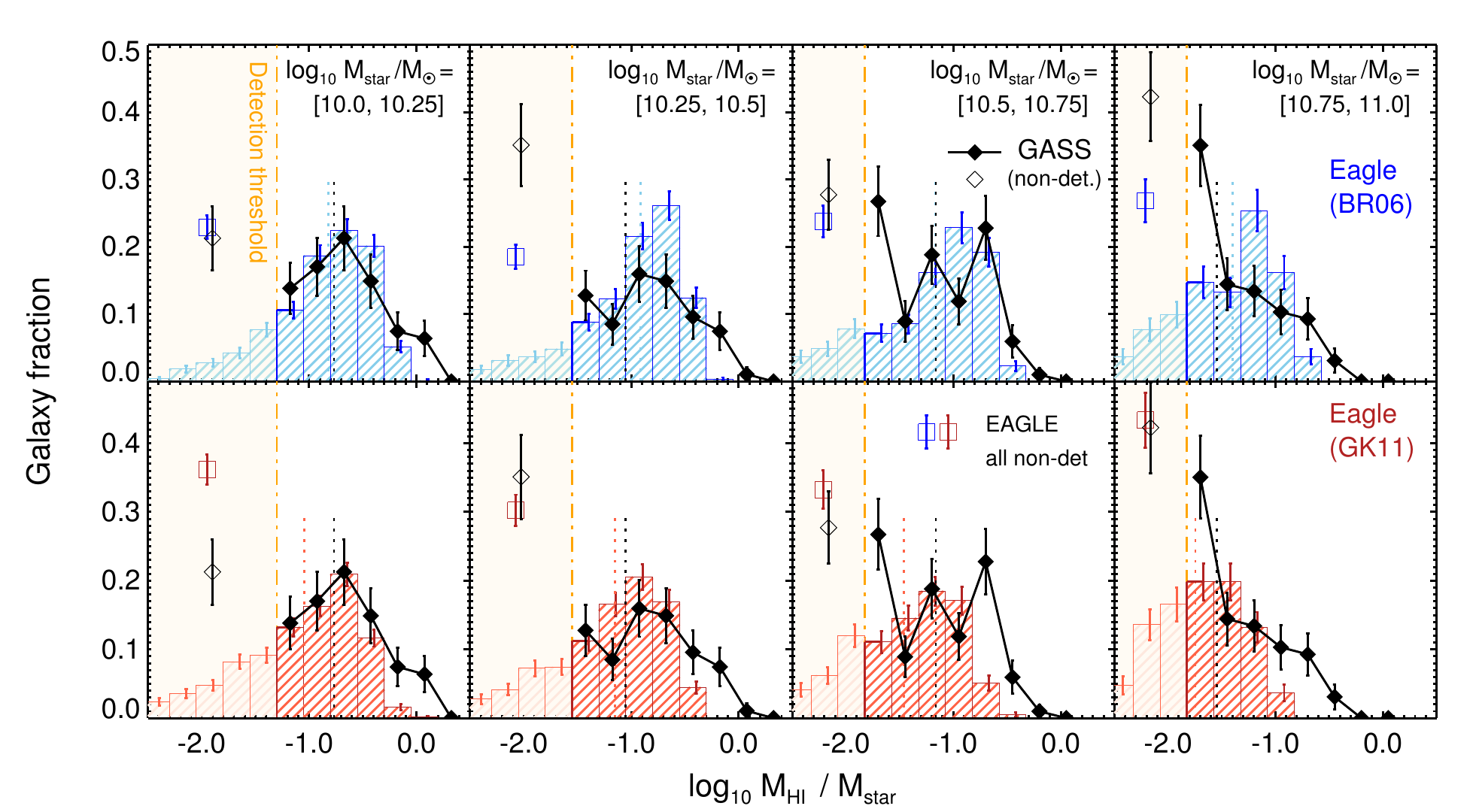}
       \caption{Comparison of the \hi{} mass of \eagle{} galaxies (blue/red histograms) with \gass{} observations (black lines); both samples include only central galaxies. In the top panel, the presence of H$_2$ in \eagle{} is accounted for with the empirical \citet{Blitz_Rosolowski_2006} pressure-law prescription, while the bottom panel shows the corresponding results from the theoretical \citet{Gnedin_Kravtsov_2011} partition formula. The shaded region on the left is below the (maximum) \gass{} detection threshold in each panel; all simulated galaxies in this regime (light blue/red) are combined into the blue/red open square for comparison to the observations. Vertical black (blue/red) dotted lines indicate the median \hi{} mass fraction of all \gass{} (\eagle) galaxies per stellar mass bin; in the third panel of the top row both lie on top of each other. Error bars show statistical Poisson uncertainties. Both H$_2$ prescriptions lead to broad agreement of the predicted \hi{} masses with observations, but the detailed match is considerably better for the BR06 H$_2$ formula (top).}
    \label{fig:gass.masshist}
  \end{figure*}

Having established that the neutral hydrogen content of \eagle{} galaxies agrees with observations, we now turn to analysing the atomic hydrogen subcomponent. Fig.~\ref{fig:gass.masshist} presents a comparison of the atomic hydrogen mass fractions, $\mhi / \mstar$ in \eagle{} with data from the \gass{} survey (\citealt{Catinella_et_al_2010}; see Section \ref{sec:surveys.gass}). We show here the distribution of $M_\hi /\mstar$ for galaxies in four narrow bins of stellar mass (individual panels, mass increases from left to right). Blue/red histograms show the distribution for simulated \eagle{} galaxies: in the top row, we adopt the empirical BR06 formula to account for the presence of H$_2$ (blue histograms), while this is achieved following the theoretical GK11 formula in the bottom row (red). In both cases, \gass{} data are represented by black lines. The vertical orange dash-dot line marks the (maximum) \gass{} detection threshold in each stellar mass bin: for consistency, we combine all galaxies with $M_\hi /\mstar$ lower than this into a single `non-detected' bin (blue/red open square / black open diamond in the shaded region on the left). 

Both \eagle{} and \gass{} show a decrease in $M_\hi / \mstar$ with increasing $\mstar$ (see also \citealt{Catinella_et_al_2010}). While both H$_2$ models (top/bottom row) lead to broad agreement with the observed distribution in shape and normalisation, the match is considerably better with the empirical H$_2$ formula of BR06 (top/blue): the median \hi{} mass fractions (vertical dotted lines) differ by $<0.2$ dex in all four bins of stellar mass and show no systematic deviation from the observed median. This level of agreement is considerably better than obtained by other recent hydrodynamic simulations (e.g.~\citealt{Aumer_et_al_2013}, whose \hi{} fractions are higher than observed by $\sim$0.5 dex; see \citealt{Wang_et_al_2014}). In contrast, the median \hi{} fraction obtained with the theoretical GK11 approach is consistently too low by $\sim0.1$ -- 0.4 dex.

It is possible that the differences between these two H$_2$ schemes are driven by inaccurate gas-phase metallicities in \eagle{} galaxies, to which the BR06 pressure law is by construction insensitive; further work is required to test whether this is indeed the case. It is also important to keep in mind that dense gas is modelled in a highly simplified way in \eagle{}, with the primary aim of circumventing numerical problems that would arise if gas were allowed to cool below $\sim$10$^4$ K at the resolution of \eagle{} (see e.g.~\citealt{Schaye_DallaVecchia_2008}). It is plausible that the BR06 and GK11 prescriptions simply reflect this imperfect ISM model in different ways, leading to different predictions about the H$_2$ fractions.

As mentioned above, \citet{Lagos_et_al_2015} obtained good agreement between the H$_2$ content of \eagle{} galaxies and observations with the GK11 prescription, which we find to yield too low \hi{} fractions. The likely reason for this apparent contradiction is that \citet{Lagos_et_al_2015} focus their analysis on the sub-sample of galaxies above the \coldgass{} detection threshold, and include both centrals and satellites: here on the other hand, we consider centrals only and calculate overall medians (from both detections and non-detections). Our results are therefore not directly comparable.

The match to \gass{} in Fig.~\ref{fig:gass.masshist} is not quite perfect even with the empirical BR06 model, however: on close inspection, the scatter in $M_\hi$ at fixed $\mstar$ is slightly smaller in \eagle{} than \gass{}, which manifests itself in a relative deficiency of non-detections (26\% vs.~42\% in the highest stellar mass bin) and very \hi-rich galaxies (a difference of -0.15 dex in the 90th percentile of $\mhi/\mstar$ in the highest stellar mass bin). The latter discrepancy is also seen with the GK11 H$_2$ model, and we confirm in Appendix \ref{sec:app.h2} that it is still present even when we ignore the presence of H$_2$ completely and assign all neutral gas as ``\hi''. Although the observational scatter may be overestimated due to uncertainties in the stellar mass measurements, we demonstrate below that a more likely cause is the presence of spuriously large \hi{} holes in the simulated galaxies. Overall, however, we can conclude that (central) \eagle{} galaxies acquire approximately realistic amounts of \hi{} by $z = 0$, with only relatively minor uncertainties introduced by the choice of model to account for the presence of H$_2$.


\section{The internal structure of \hi{} in simulated galaxies}
\label{sec:profiles}

We now investigate the internal distribution of atomic hydrogen in the simulated galaxies. Even in light of the good match between total \hi{} masses in \eagle{} and observations as demonstrated above, there is no guarantee that the former is modelled in an equally realistic way: many previous hydrodynamical simulations have suffered from `overcooling' which leads to an artificially enhanced gas density in the central region, especially in massive galaxies (e.g.~\citealt{McCarthy_et_al_2012b}; see also \citealt{Crain_et_al_2015}). In combination with a deficit of gas in the outskirts, the total \hi{} mass in simulated galaxies could agree with observations even in this case.

\subsection{Visual inspection of \hi{} morphologies}
\label{sec:morphology}
\begin{figure*}
  \centering
    \includegraphics[width=2.1\columnwidth]{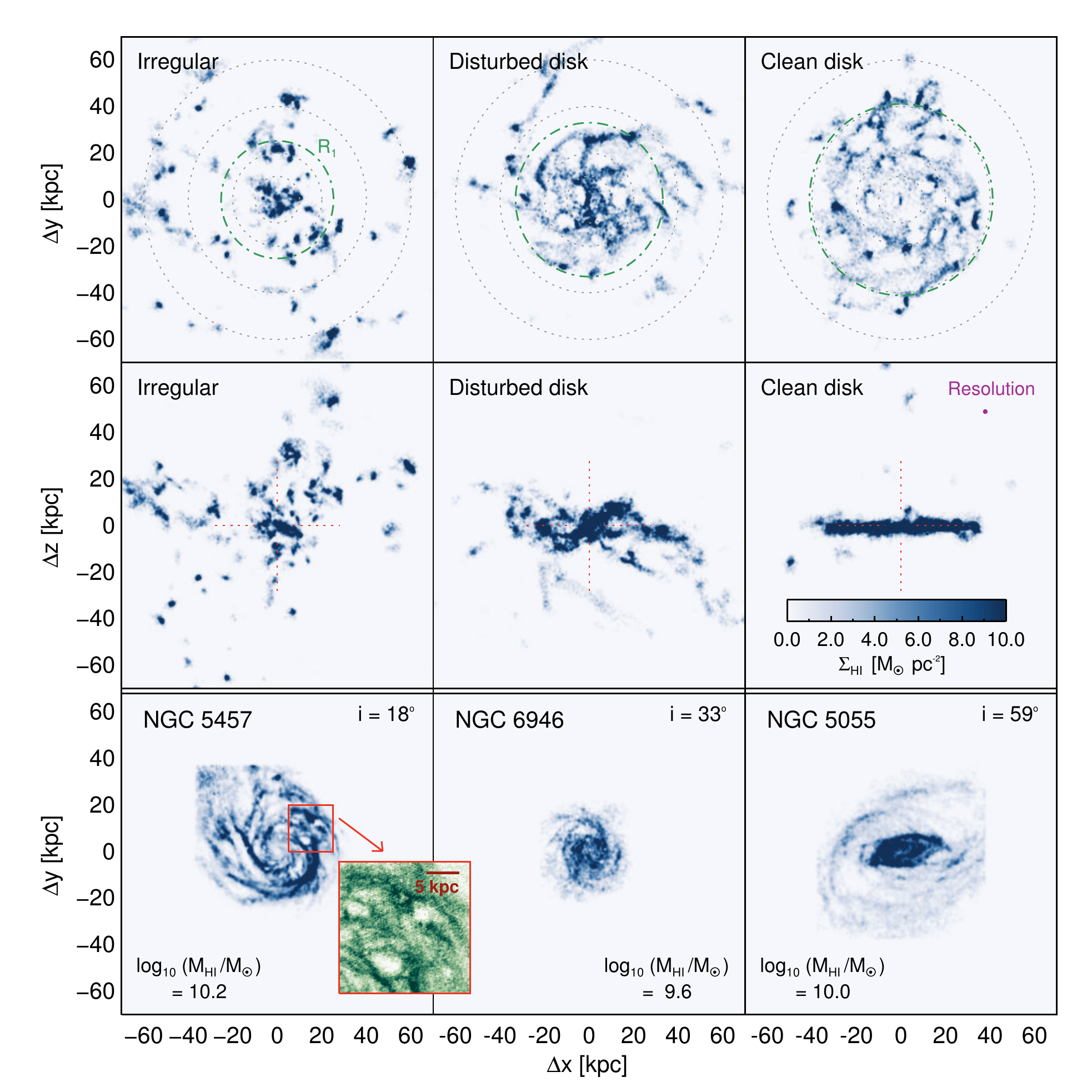}
       \caption{Top and middle row: Three typical examples of \eagle{} galaxies with different \hi{} morphologies, but similar \hi{} mass $\log_{10} \,(\mhi/\msun) = [9.9, 10.1]$: Irregular (\textbf{left}), Disturbed disk (\textbf{middle}) and Clean disk (\textbf{right}). Face-on images are shown in the top row, edge-on equivalents below. For comparison, the bottom row shows three observed \hi{} images from \emph{THINGS} \citep{Walter_et_al_2008} at the same physical scale; these do not correspond to the same morphology categories as the simulated galaxies above. All images use the same linear scaling, as given in the middle-right panel, and are Gaussian-smoothed to a (FWHM) resolution of 1 kpc (purple circle in the middle-right panel). The green dash-dot rings in the top row show the characteristic radius $R_1$ for each galaxy (see Section \ref{sec:mass-size}); the grey dotted circles indicate radii of 10, 20, 40, and 60 kpc, respectively. The red dotted ``cross-hairs'' in the middle row indicate the best-fit \hi{} disk plane and axis. The inset in the bottom row shows a 20 x 20 kpc zoom-in of NGC 5457, revealing \hi{} holes similar to what is seen in \eagle{} but on a smaller scale.}
    \label{fig:bd.visual_images}
  \end{figure*}

As a first qualitative step, we have created mock \hi{} images of all simulated galaxies satisfying our selection criteria, by assigning the \hi{} mass of particles within a line-of-sight interval of [-70, 70] kpc relative to the galaxy centre to an $x \times y$ grid with pixels of 0.5 kpc, and smoothing with a Gaussian FWHM of 1 kpc. For simplicity, we only do this with the empirical BR06 H$_2$ correction. All images were then inspected visually, and the galaxies assigned to one of three broad morphological categories: (a) `Irregular' (no disk-like structure), (b) `Disturbed \hi{} disks' (which are not flat when edge-on, and instead show e.g. prominent warps), and (c) `Clean \hi{} disks'. Their relative abundances will be discussed in Section \ref{sec:morph.abundances} below. While this is inevitably a subjective classification, it can still offer valuable insight into the \hi{} structure that may not be apparent from a simple quantitative analysis. A typical example galaxy from each category is shown in the first two rows of Fig.~\ref{fig:bd.visual_images}; each has similar total \hi{} mass ($\log_{10} \mhi/\msun = [9.9, 10.1]$) but rather different appearance. Galaxies have been rotated to face-on in the top row, and to edge-on in the middle; the disk plane is defined to be perpendicular to the angular momentum axis of all \hi{} within a spherical 50 kpc aperture which corresponds roughly to the edge of the largest \hi{} disks in our sample (see Fig.~\ref{fig:bd_r1} below). The scaling is linear from 0 to 10 $\msun$ pc$^{-2}$ with darker shades of blue indicating denser gas; the smoothing scale of 1 kpc is indicated with a purple circle in the middle-right panel.

\begin{figure*}
  \centering
    \includegraphics[width=2.1\columnwidth]{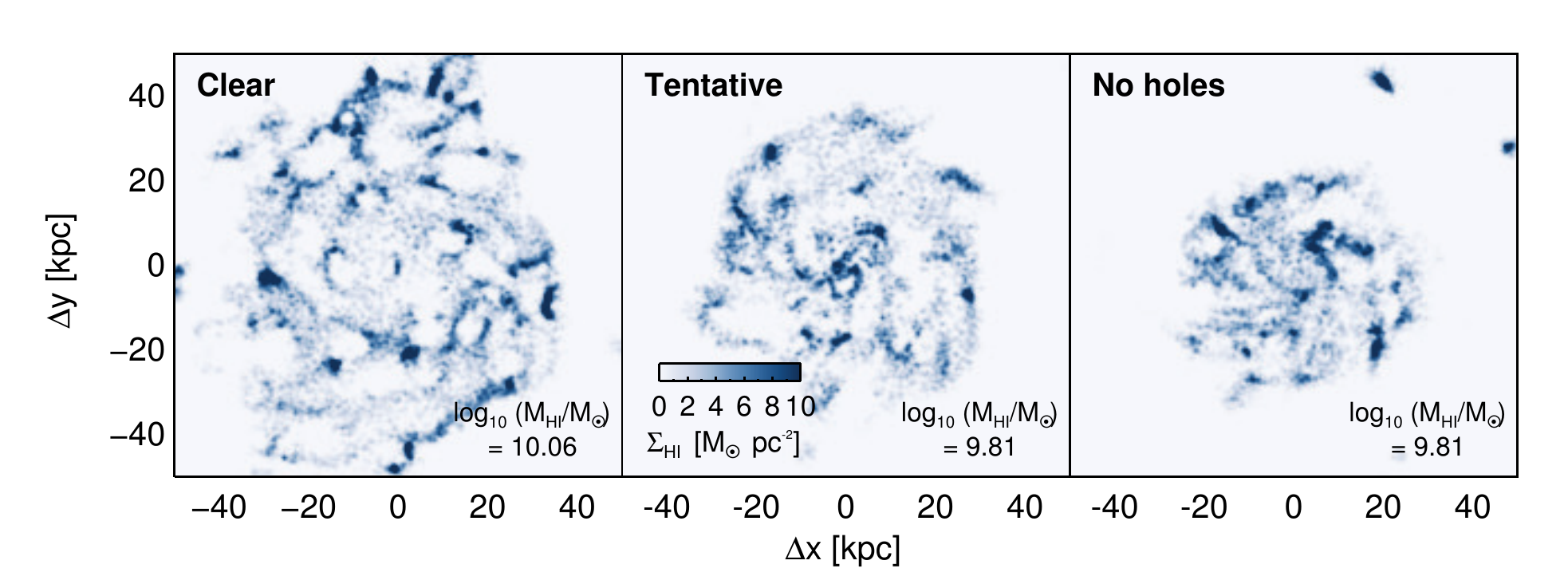}
       \caption{Examples of simulated face-on galaxy \hi{} images with clear (\textbf{left}), tentative (\textbf{middle}) and no (\textbf{right}) \hi{} holes as found by visual inspection; all of them have similar \hi{} mass. The scaling is the same as in Fig.~\ref{fig:bd.visual_images}. Note that even the ``no holes'' galaxy in the right panel shows some hole-like structures, but these are much smaller than those seen in the left panel and therefore not classified as ``holes'' here.}
      \label{fig:bd.holetypes}
  \end{figure*}

`Irregular' galaxies in particular (left-hand panels) typically contain a large number of \hi{} `blobs' that are typically representing gas in the process of accreting onto the galaxy. Although the `disk' galaxies (middle and right columns) have their \hi{} predominantly in a more or less thin disk, many of these also show pronounced substructure with dense clumps as well as large ($\sim$10--20 kpc) \hi{} holes. To illustrate the varying degree to which the latter are apparent in different galaxies, we show in Fig.~\ref{fig:bd.holetypes} mock face-on \hi{} images of three galaxies: one with clearly visible holes (left), one with only tentative hole identifications (middle), and one without visible large holes (right).

We have verified that these holes are also found in the total gas density maps, so they are not an artefact of our \hi{} modelling (i.e. they are not regions where most of the neutral gas is H$_2$). From inspection of the high time-resolution ``snipshot'' outputs in \eagle{} (see \citealt{Schaye_et_al_2015}), they form rapidly and can reach sizes of $\sim$10 kpc within only 20 Myr. This suggests that they are the result of heating events associated with star formation that are (individually) orders of magnitude more energetic than supernova explosions in the real Universe as a result of the limited resolution of \eagle. Their detailed formation and survival is almost certainly more complex, however: as we show below, a clear correlation between star formation and occurrence of holes is not observed in the \eagle{} galaxies.

For comparison, the bottom row of Fig.~\ref{fig:bd.visual_images} displays three observed \hi{} maps of nearby spiral galaxies of comparable $\mhi$ --- NGC 5457, NGC 6946, and NGC 5055 --- from The \hi{} Nearby Galaxies Survey (\emph{THINGS}; \citealt{Walter_et_al_2008}) smoothed to the same spatial resolution as our simulated maps (1 kpc). These clearly look different from even the simulated `clean disk' galaxies and have a much smoother but also more intricate structure (see also \citealt{Braun_et_al_2009}) with clear spiral arms. This difference in appearance can be attributed to the imperfect modelling of the ISM in \eagle, which does not explicitly include a cold phase, and must therefore impose a pressure floor for high density gas \citep{Schaye_DallaVecchia_2008, Schaye_et_al_2015}. In addition, the observed galaxies do not show \hi{} holes comparable to those in the simulation, although such features do occur on smaller scales \citep{Boomsma_et_al_2008}: the small inset in the bottom row shows a zoom of a 20 x 20 kpc region of the almost face-on galaxy NGC 5457 which clearly shows numerous holes up to scales of $\sim$5 kpc (note that the inset uses a square-root scaling for improved clarity, and is therefore shown in a different colour). Rather than the \emph{existence} of \hi{} holes being an artefact of the simulation, it is rather their size and hence covering fraction in the disk that is in tension with observations.

Another apparent disagreement between the simulations and observations is the thickness of the \hi{} disks: as the middle-right panel shows, even a `clean disk' extends several kpc in the vertical direction. The exponential scale-height of this disk is $\sim$1.5 kpc, several times larger than the values indicated by observations of \hi{} in the Milky Way \citep{Dickey_Lockman_1990}. We note, however, that this is only a factor of $\sim$2 larger than the gravitational softening length of the simulation, which may largely explain this discrepancy. Another plausible contributor is the imposed temperature floor of $\sim$10$^4$ K, which corresponds to a Jeans length of $\sim$1 kpc. It is conceivable that the artificial thickening of the disk and the unrealistically large size of \hi{} holes noted above are in fact related: a thicker disk implies a lower volume density and hence less mass swept up per unit distance. Despite these differences in detail, we show below that the azimuthally averaged distribution of \hi{} in our simulated galaxies agrees quantitatively with observations, both in terms of the \hi{} sizes and surface density profiles.

\subsection{Correlations of \hi{} morphology with other galaxy properties}
\label{sec:morph.abundances}
\begin{figure*}
  \centering
    \includegraphics[width=2.1\columnwidth]{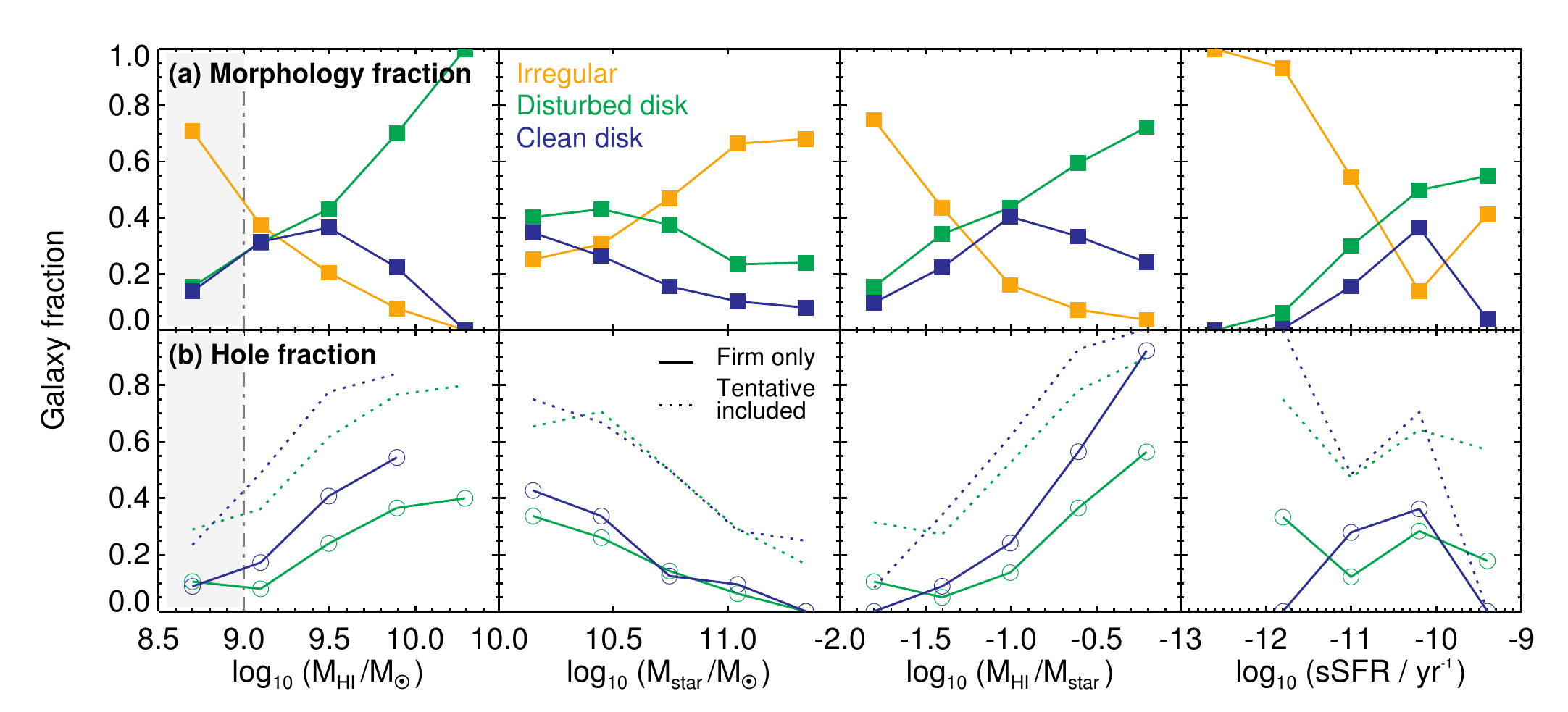}
       \caption{Fraction of galaxies with different \hi{} morphologies (\textbf{top}), and with visible \hi{} holes (\textbf{bottom}, dotted lines include tentative identifications). From left to right, the individual columns show the fractions as a function of \hi{} mass, stellar mass, \hi{} mass fraction and specific star formation rate (sSFR). The fraction of disturbed disks increases strongly with \hi{} mass, in qualitative agreement with observations; the same is true for the fraction of galaxies with visible \hi{} holes. Trends are also seen with the other galaxy parameters, as discussed in the text. The grey region in the left-hand column is affected by our imposed stellar mass limit.}
    \label{fig:bd.morphtypes}
  \end{figure*}

The relative abundance of each morphological type as a function of, respectively, the total \hi{} mass, stellar mass, \hi{} mass fraction and specific star formation rate (sSFR) is shown in the top panels of Fig.~\ref{fig:bd.morphtypes}. \hi{} morphology correlates strongly with all four of these parameters: irregular distributions are most common at low \hi{} mass, high $\mstar$ and low sSFR, whereas the fraction of disks is highest in the opposite regimes. Interestingly, the fraction of \emph{clean} disks (blue line in Fig.~\ref{fig:bd.morphtypes}) shows no such simple behaviour: they are most common at intermediate $\mhi \approx 10^{9.4}\, \msun$, while most galaxies at the high $\mhi$ end show a `disturbed disk' morphology. As \hi{} content and sSFR are correlated, it is not surprising to see similar trends with the latter (rightmost panel): the fraction of clean disks drops sharply for the most actively star forming galaxies. 

Note that the left-hand column in particular (trends with $\mhi$) suffers from incompleteness due to our imposed stellar mass threshold of $\mstar \geq 10^{10}\,\msun$: there is a large population of less massive galaxies, many of which with $\mhi$ high enough to fall within the range plotted here. With a median $\mhi/\mstar = 0.1$ at $\mstar = 10^{10}\, \msun$ (see Fig.~\ref{fig:gass.masshist}), this mostly affects the range $\mhi \lesssim 10^9 \msun$ which is therefore shaded grey in Fig.~\ref{fig:bd.morphtypes}. This should be kept in mind when comparing to \hi{} limited surveys, but insofar as only \emph{massive} galaxies are concerned ($\mstar \geq 10^{10}\, \msun$), it does not affect our results.

The bottom panels of Fig.~\ref{fig:bd.morphtypes} show the fraction of galaxies in the two `disk' categories with one or more \hi{} holes. They are generally more common in clean than disturbed disks and at higher $\mhi$ (e.g.~55 per cent at $\mhi \approx 10^{10}\, \msun$) and lower $\mstar$. Including tentative hole identifications (dotted lines), their occurrence increases by a factor of $\sim$2 to $\sim$80 per cent at both the low-$\mstar$ and high-$\mhi$ ends. Perhaps surprisingly, the hole fraction shows no clear increase with increasing sSFR, and when tentative detections are included, there is a clear increase towards \emph{lower} sSFR, at least in clean disks (blue). Even though we here show current sSFR --- which may well already have been lowered by the presence of low-density holes --- we have tested for correlation between holes and star formation in the recent past (as well as total star formation and SFR density), which yields a similar result. This suggests a complex connection between star formation (and the associated feedback) and the occurrence of holes. It is possible, for example, that disk instabilities can prolong the lifetime of holes and therefore make them a more prominent feature in \hi-rich galaxies (see also \citealt{Mitchell_et_al_2012, Agertz_et_al_2015}). 

In summary, a large fraction of our simulated galaxies (64 per cent) show a disc morphology of their \hi{} content (particularly at high $\mhi$), but nearly two thirds of these are evidently disturbed. All \hi{} discs are a factor of several too thick and lack the intricate spiral structure seen in observed \hi{} maps, which is a direct consequence of the simplified ISM modelling in \eagle{}. Particularly at high $\mhi$ and low $\mstar$, many galaxies furthermore show \hi{} holes that are larger than what is observed (43 and 19 per cent of all central galaxies, respectively, with and without tentative hole identifications).

\subsection{\hi{} size--mass relation}
\label{sec:mass-size}
As a simple one-parameter proxy for the internal gas distribution, we next investigate the `characteristic' size of the \hi{} discs, which we define as the radius $R_1$ at which the azimuthally averaged surface mass density drops below 1 $\msun$ pc$^{-2}$; this definition is commonly encountered in the observational literature (e.g.~\citealt{Broeils_vanWoerden_1994,Wang_et_al_2013,Wang_et_al_2014}). To find $R_1$, we align all galaxies to face-on (see above) and generate two-dimensional \hi{} surface density maps $\Sigma_\hi$ with a pixel size of 0.5 kpc, integrating along the line of sight over the range [-70, +70] kpc. We compare our data to the mass--size relation of \citet{Broeils_Rhee_1997}; their galaxies were observed at the \emph{Westerbork Synthesis Radio Telescope} (WSRT) with a beam size of $\sim$12$^{\prime\prime}$ and are at a median distance of $\sim$35 Mpc \citep{Broeils_vanWoerden_1994} which corresponds to a physical resolution of $\sim$2 kpc. Note that this relation has recently been verified and extended to very \hi{}-rich galaxies by \citet{Wang_et_al_2013}. We convolve our \hi{} maps with a Gaussian kernel of FWHM = 2 kpc, and extract radial profiles in a set of (circular) annuli of width 2.5 kpc. $R_1$ is then determined by interpolating linearly between the outermost bin with density above the threshold of $1 \msun\, \text{pc}^{-2}$, and the one beyond this\footnote{We note that this procedure is not strictly self-consistent, because such an interpolation yields, in general, a cumulative mass profile that differs from the true profile. However, we have experimented with more elaborate methods such as linear or quadratic spline fits, or narrower profile bins. None of these alternatives differ substantially from the simple method adopted here.}. In the top row of Fig.~\ref{fig:bd.visual_images}, $R_1$ is shown as a dark green dash-dot circle, and coincides approximately with what one would visually identify as the `outer edge' of the \hi{} disk.

\begin{figure}
  \centering
    \includegraphics[width=\columnwidth]{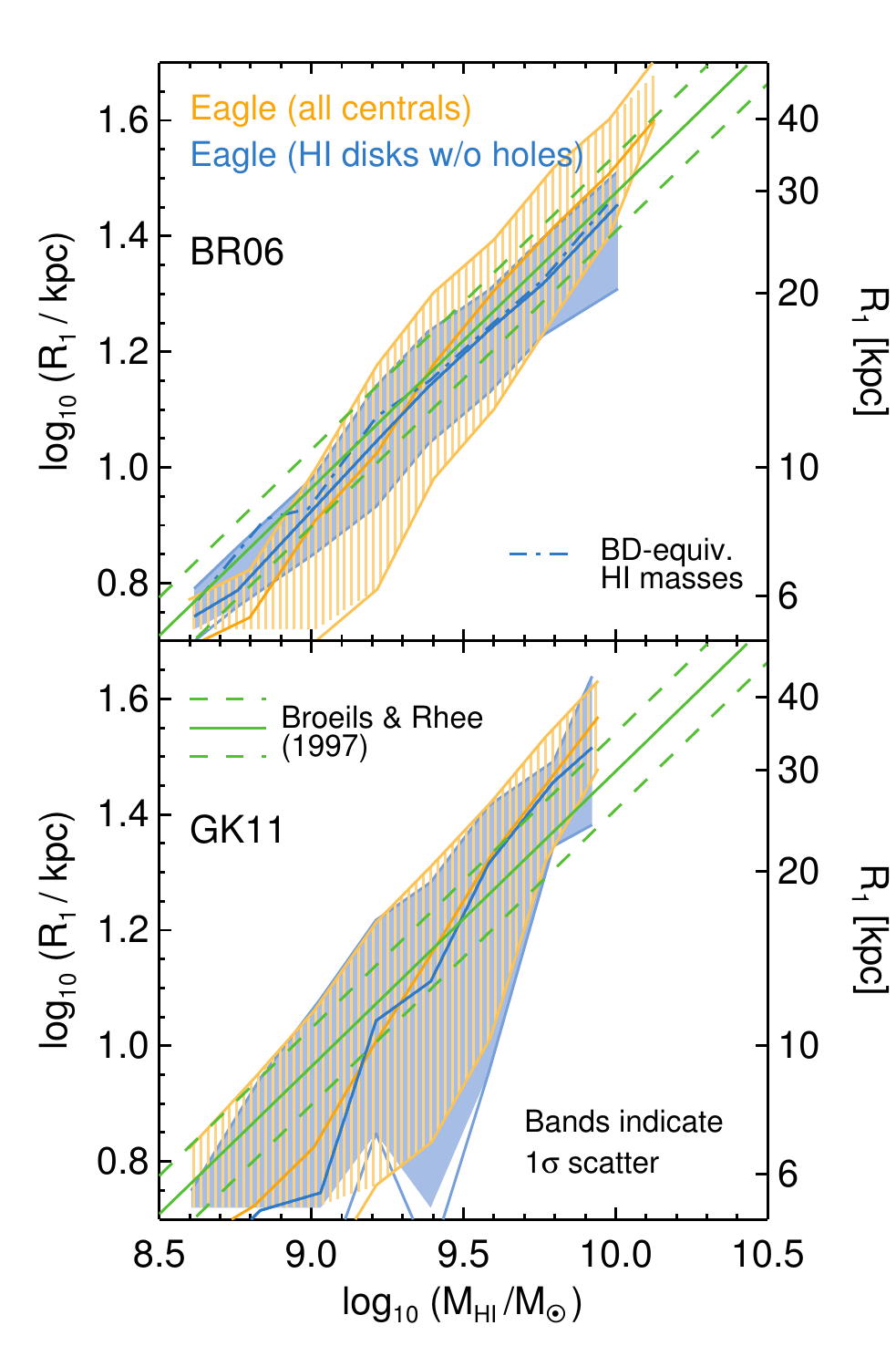}
       \caption{The \hi{} mass--size relation for \eagle{} galaxies (blue/orange bands), compared to the observational data from \citet{Broeils_Rhee_1997} (green lines). Orange includes all (central) \eagle{} galaxies whereas blue only includes those with a visually confirmed \hi{} disk that did not show prominent \hi{} holes. The \textbf{top} panel shows results with the empirical \citet{Blitz_Rosolowski_2006} H$_2$ model, whereas the theoretical \citet{Gnedin_Kravtsov_2011} approach is used in the \textbf{bottom} panel. For the former, the agreement of the median trends is already quite good in the full sample, and even better for the ``clean'' subset (blue) over more than an order of magnitude in \hi{} mass, especially when using ``\bluedisk{}'' equivalent \hi{} masses (dash-dot line; see text). The theoretical model (bottom) predicts an \hi{} mass--size relation that is significantly too steep.}
    \label{fig:bd_r1}
  \end{figure}

In Fig.~\ref{fig:bd_r1}, we show the resulting relation between $R_1$ and $M_\hi$ for \eagle{} galaxies. As before, we select only central galaxies\footnote{We have verified that the result is virtually unchanged when potentially interacting galaxies are excluded, i.e.~those where a neighbouring galaxy within 150 kpc has a stellar mass exceeding one tenth of its own.}; we show results for both the empirical BR06 (top) and theoretical GK11 prescription (bottom) to account for H$_2$. The running median and $1\sigma$ scatter (defined as the 15.9/84.1 percentile) of these is shown in orange. To test the influence of \hi{} morphology, we also show (in blue) the size--mass relation for only the subset of simulated galaxies that have been visually classified as containing a well-aligned \hi{} disk without prominent holes\footnote{Dropping the requirement for the galaxies to have an \hi{} disk makes no noticeable difference, but the \citet{Broeils_Rhee_1997} sample includes only spiral and irregular galaxies (by optical morphology).}. The green solid line represents the best fit of \citet{Broeils_Rhee_1997}, with $1\sigma$ standard deviation indicated by the green dashed lines (note that our plot has the $x$- and $y$-axes swapped relative to their figure 4, and that they show $D_1 \equiv 2R_1$ instead).

In general, the \eagle{} galaxies follow the observed relation quite well. The full sample (orange) has a slope that is somewhat steeper than observed with both H$_2$ recipes (0.59 vs.~0.51, i.e.~a 16 per cent difference with BR06, and 52 per cent with GK11). With the empirical BR06 method (top), the `disk-only, no holes' distribution (blue) agrees with the observations to better than 0.1 dex over the full range of $\mhi$ that we probe here, $\log_{10}\,  (\mhi/\msun) = [8.5, 10.0]$. The agreement gets better still when we calculate the \hi{} masses in analogy to the \bluedisk{} survey (see below), the observations for which were also conducted at the WSRT: this reduces $\mhi$ slightly at the low-$\mhi$ end (see top panel of Fig.~\ref{fig:app.mhicomparison}) and therefore improves the agreement of \eagle{} with the \citet{Broeils_Rhee_1997} relation to $< 0.05$ dex. We note, however, that \citet{Broeils_vanWoerden_1994} find close agreement between the interferometry-derived \hi{} masses used by \citet{Broeils_Rhee_1997} and single-dish measurements, so it is not clear whether this is indeed a more fair comparison to the data. We also note that the scatter in the overall \eagle{} sample is somewhat too large, by a factor of $\sim 2$. Again, the agreement here is better in the sample that excludes galaxies with prominent \hi{} holes (blue). With the GK11 $H_2$ formula (bottom), excluding holes makes no appreciable difference, and the \hi{} size--mass relation remains steeper and broader than observed.

\subsection{Density profiles for \hi-rich and ``normal'' galaxies}
\label{sec:bd.profiles}
A more detailed quantitative test of the \hi{} structure is to compare the radial (surface) density profiles to observations. A particularly interesting question in this respect is how atypically \hi-rich galaxies -- which have likely been particularly efficient at accreting \hi{} recently -- compare to those with average \hi{} content. Motivated by this consideration, the \bluedisk{} survey \citep{Wang_et_al_2013} has recently observed a set of 25 galaxies expected to be \hi-rich, and a similar number of `control' galaxies, generating resolved \hi{} maps with a resolution of $\sim$10 kpc. One key discovery of this study has been that the \hi{} surface density profiles of \emph{all} galaxies, both \hi-rich and normal, follow a `universal' shape as long as they are normalised by the characteristic \hi{} disc size $R_1$ \citep{Wang_et_al_2014}. We will now compare the \eagle{} galaxies to these observations. For simplicity, we focus first on results obtained with the empirical \citet{Blitz_Rosolowski_2006} H$_2$ model which, as shown above, leads to total \hi{} masses in good agreement with observations. Profiles obtained with the theoretical H$_2$ formula of \citet{Gnedin_Kravtsov_2011} will be presented in Section \ref{sec:difference.reason} below.

\subsubsection{Sample definitions}
From our 2083 central \eagle{} galaxies with $\log_{10}\, (\mstar/\msun) = [10.0, 11.0]$ -- the same range as in \bluedisk{} -- we select those with a Petrosian half-light radius $R_{50,z} \geq 3$ kpc. This radius is defined as enclosing 50 per cent of the Petrosian flux in the SDSS z-band\footnote{As detailed in Appendix \ref{sec:app.bdsel-size}, the radii $R_{50,z}$ obtained for our simulated galaxies are systematically too large compared to observations from SDSS; this is investigated in more detail by \citet{Furlong_et_al_2015b}. For our purpose, we simply re-scale the distribution of $R_{50,z}$ to enforce a match to the observational data. This ad-hoc fix does not invalidate our results below, because we are only concerned with relative size comparisons, and only use the stellar sizes to select the overall sample to compare to \bluedisk{}.} obtained from stellar population synthesis (SPS) modelling \citep{Trayford_et_al_2015}. As we show in Appendix \ref{sec:app.bdsel-size}, this simple size cut approximately reproduces the more complex original sample selection in the \bluedisk{} survey \citep{Wang_et_al_2013}. We have verified that the profiles shown below are insensitive to the exact value of the size cut, and are actually almost unchanged when all galaxies are included, regardless of size.

This simulated sample with 607 members is then divided into `\hi-rich' galaxies with $\mhi \geq 10^{9.8}\, \msun$, and a `control' sample with $10^{9.1}\, \msun \leq \mhi \leq 10^{9.8}\, \msun$. As opposed to an -- equally plausible -- set of cuts in $\mhi/\mstar$, these limits approximately correspond to the sample division in \bluedisk{} (see \citealt{Wang_et_al_2013} and Appendix \ref{sec:app.bdsel-rich}). For consistency with the observations, we calculate \hi{} masses here in a ``\bluedisk-equivalent'' fashion \citep{Wang_et_al_2013}: a two-dimensional \hi{} surface density map with pixel size 0.5 kpc was created (with the simulation $z$-coordinate as the line of sight, i.e.~random galaxy orientations) and then smoothed with an elliptical Gaussian of FWHM = 14 (9) kpc major (minor) axes. This corresponds approximately to the WSRT beam size at the median redshift of the \bluedisk{} galaxies ($\tilde{z} \approx 0.027$). From these maps, we then sum over all pixels with $\Sigma_\hi$ above the median \bluedisk{} detection threshold of $4.6 \times 10^{19}$ atoms cm$^{-2}$ ($= 0.37\, \msun$ pc$^{-2}$ in \hi) to obtain the total \hi{} mass of the galaxy. In Appendix \ref{sec:app.aperture} we show that the resulting $\mhi{}$ are typically less than 0.1 dex below the ``\gass-equivalent'' mass for $\mhi \geq 10^{9.1} \msun$.

Recall from Section \ref{sec:gass} that \eagle{} has a deficiency of galaxies at the \hi-rich end. This is unfortunate for the present purpose, because it means that our simulated \hi-rich galaxies are typically not quite as extreme as those in the corresponding \bluedisk{} sample. However, the two populations are still clearly different and enable us to test the universality of the \hi{} density profile: The 406 simulated `control' galaxies have a median $\log_{10} (\mhi/\msun) =  9.61$, whereas the 133 `\hi-rich' counterparts\footnote{There are some additional galaxies with $\mhi < 10^{9.1} \msun$, which are not included in either sample.} have a median $\log_{10} (\mhi/\msun) = 9.93$. For comparison, the Bluedisk \hi-rich galaxies have a median $\log_{10} (\mhi/\msun) = 10.09$ ($N = 23$) and the control sample $\log_{10} (\mhi/\msun) = 9.59$ ($N = 18$). The difference between the two \eagle{} samples (0.32 dex) is therefore approximately two thirds of that in Bluedisk (0.5 dex).

\subsubsection{Density profile comparison}
\label{sec:profiles.comparison}
In \citet{Wang_et_al_2014}, density profiles were extracted using elliptical annuli with orientation and axis ratio ($b/a$) taken from the best-fit ellipse to the stellar r-band light, and then multiplied by a factor of $b/a \approx \cos(\theta)$ to correct the profiles to face-on. For the analysis of our simulated galaxies, we simply rotate them to face-on by aligning the angular momentum axis of the \hi{} in the central 50 kpc with the line of sight (as in Fig.~\ref{fig:bd.visual_images}). For each galaxy, an \hi{} image was then created as described above, and the surface density profile extracted with 20 equally spaced bins from 0 to $2R_1$ (recall that $R_1$ is defined as the radius at which the \hi{} surface density drops to 1 $\msun$ pc$^{-2}$). Galaxies were then median-stacked to obtain the average profile for the \hi-rich and control galaxies. The same procedure was applied to the Bluedisk profiles from \citet{Wang_et_al_2014}.

\begin{figure}
  \centering
    \includegraphics[width=\columnwidth]{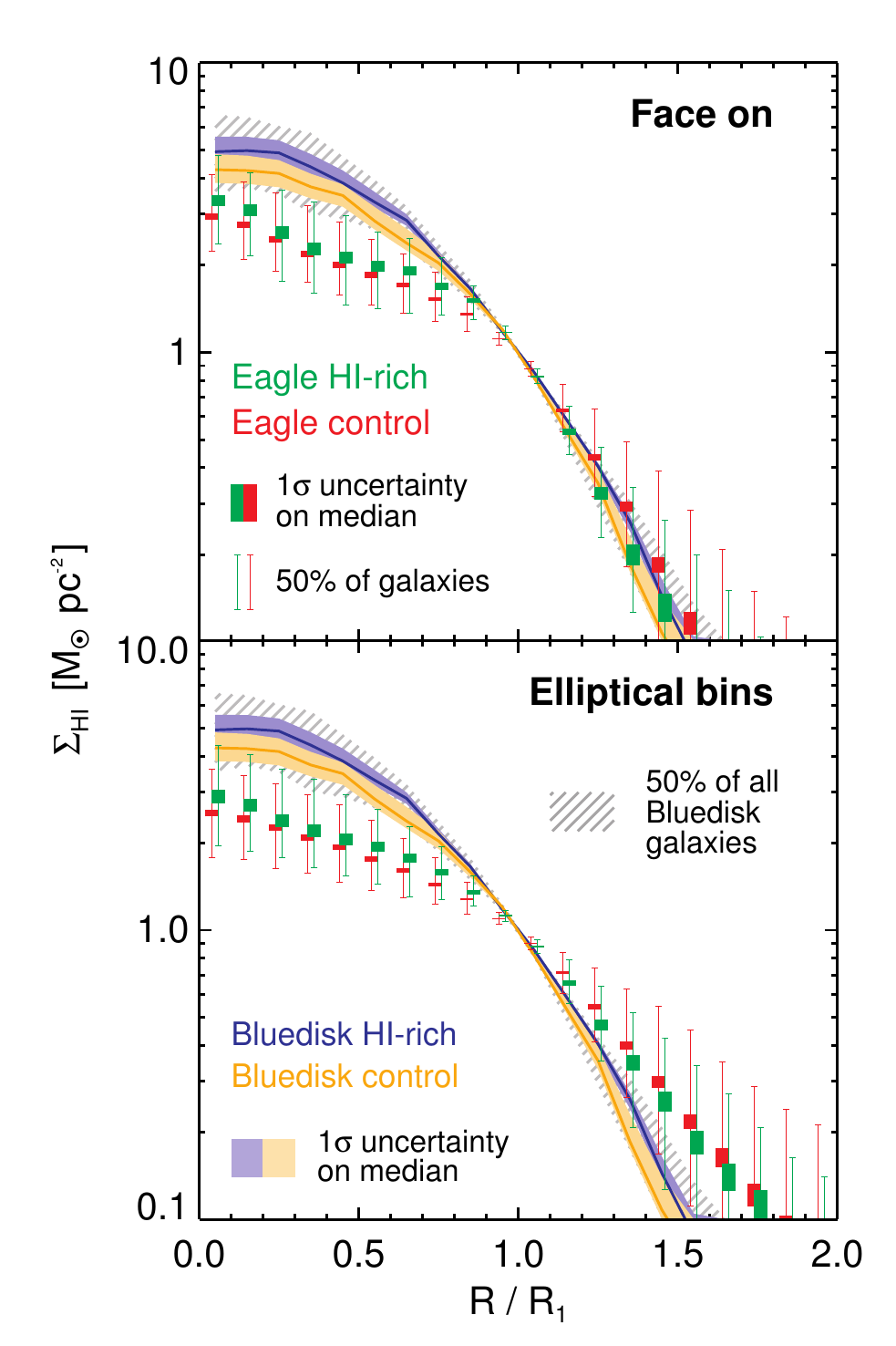}
           \caption{The scaled \hi{} surface density profiles for galaxies in the \eagle{} simulation (red/green rectangles) and in the Bluedisk survey (yellow/blue shaded bands). Galaxies are split into ``\hi-rich'' and ``control'' samples based on their \hi{} mass, as explained in the text. The \bluedisk{} profiles are identical in both panels, but different methods are applied to \eagle{}: \textbf{Top:} galaxies are rotated to face-on. \textbf{Bottom:} profiles are extracted in random orientation in elliptical bins with position angle and axis ratio determined from the stellar r-band light (as in the \bluedisk{} analysis). In agreement with observations, both methods yield similar profiles for simulated \hi-rich and control galaxies. However, the \eagle{} profiles deviate from the observations in the central region, and with the `elliptical bin' inclination correction (bottom), they are also too shallow in the outskirts.}
          
    \label{fig:bd.profiles}
  \end{figure}

The result is shown in the top panel of Fig.~\ref{fig:bd.profiles}. Yellow and blue lines trace the median profiles for \hi-rich and control galaxies in \bluedisk{}. As in Fig.~\ref{fig:neutralgas}, shaded bands indicate the statistical $1\sigma$ uncertainty on the median, i.e.~they extend from $\sigma_\text{low}$ to $\sigma_\text{high}$ where $\sigma_\text{low (high)} = \tilde{\Sigma}_\hi + (P_\text{15.9 (84.1)}-\tilde{\Sigma}_\hi) / \sqrt{N}$; $\tilde{\Sigma}_\hi$ here denotes the median and $P_n$ the $n$th percentile of the $\Sigma_\hi$ distribution in a bin with $N$ galaxies. The simulated profiles are shown with green (red) boxes for \hi-rich (control) galaxies whose vertical extent gives the $1\sigma$ uncertainty on the median calculated in the same way; they are small due to the comparatively large sample size. In addition, the galaxy-to-galaxy scatter is indicated by thin error bars for \eagle{} and by the grey hatched region for \bluedisk, both of which show the interval occupied by 50 per cent of galaxies (i.e.~$P_{\,25}$ to $P_{\,75}$).

In the outer region ($R \gtrsim R_1$), simulated and observed profiles generally agree to within the statistical uncertainties. Both simulated control and \hi-rich galaxies follow an exponential profile (straight line in the log-linear plot), although the gradient is slightly steeper for the \hi-rich galaxies (green, discrepancy at $\sim$2$\sigma$ level). The observed galaxy profiles show an approximately equal, but opposite difference (steeper profiles for control galaxies), although the small number of \bluedisk{} galaxies means that this difference is not statistically significant. 

In the inner parts ($R \lesssim R_1$), there is a much more pronounced discrepancy between simulations and observations: the former ``under-cut'' the observed profile. Interestingly, the simulated \hi-rich and control profiles are still almost identical, to within $<0.1$ dex (with a minor excess, significant at $\sim$2$\sigma$, in \hi-rich galaxies, as in observations). To test whether this is a result of the slightly different analysis for the simulated and observed galaxies, we have reproduced the \bluedisk{} analysis on our simulated galaxies exactly (i.e.~we extracted the profiles in elliptical rings with position angle and axis ratio given by the stellar r-band, and multiplied with a correction factor of $b/a$), the result of which is shown in the bottom panel of Fig.~\ref{fig:bd.profiles}. However, this fails to ameliorate the tension in the central region, and adds another disagreement in the outer parts, where the simulated profiles are now far too shallow; we shall return to this shortly.

\begin{figure}
  \centering
    \includegraphics[width=\columnwidth]{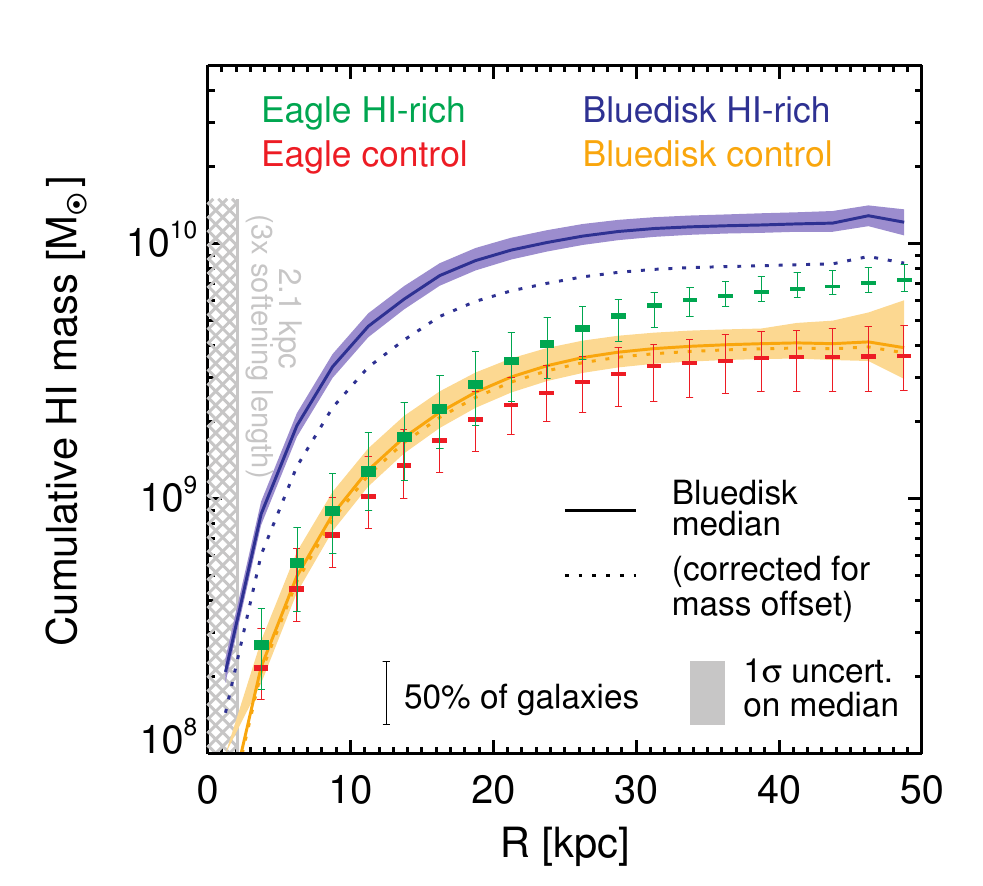}
           \caption{Cumulative \hi{} mass profiles for \eagle{} galaxies (green/red) compared to observations from \bluedisk{} (blue/yellow); different colours denote control/\hi-rich galaxies as in Fig.~\ref{fig:bd.profiles}, but here the $x$-axis is not normalised by $R_1$. For \bluedisk{}, dotted lines show the profiles corrected for the offset in total \hi{} mass in each sample relative to \eagle{} (see text). Bands and thick boxes indicate the $1\sigma$ uncertainty on the median. There is a clear lack of \hi{} in the central region ($R \lesssim$ 20 kpc) in \eagle, which manifests itself in the lower-than-observed \hi{} masses of \hi-rich galaxies (blue/green). For the `control' sample, the \hi{} deficit in the centre is largely compensated by the slightly too shallow decline of the profile in the outer parts (yellow/red).}
          
    \label{fig:bd.cummass}
  \end{figure}

It is worth keeping in mind that, due to the scaling of the $x$-axis in Fig.~\ref{fig:bd.profiles} by $R_1$, one cannot directly infer the actual mass distribution from the profiles. We therefore explicitly show a comparison between the cumulative mass profiles in \eagle{} and \bluedisk{} in Fig.~\ref{fig:bd.cummass}; the symbols have the same meaning as in Fig.~\ref{fig:bd.profiles} with the addition of the grey hatched region denoting the (small) zone influenced by resolution effects ($R \leq\, 3 \epsilon$, the gravitational softening length). Note that the $x$-axis here shows the actual galacto-centric radius $R$ in kpc, and is not normalised by $R_1$. As expected, simulated galaxies show a deficit of mass in the inner region ($R \lesssim 30$ kpc). However, for the control galaxies (red) this is almost completely compensated by the outer parts, where the surface density profile is slightly too shallow. In the \hi-rich sample, on the other hand, the central deficit manifests itself in a lower total \hi{} mass in simulated galaxies, as already seen in Fig.~\ref{fig:gass.masshist}. To confirm that this interpretation also holds quantitatively, the dashed lines in Fig.~\ref{fig:bd.cummass} show the \bluedisk{} profiles re-scaled by the ratio of median total $\mhi$ in \eagle{} and \bluedisk, 0.95 and 0.69 respectively for the \hi-rich and control samples. As expected, this shows much better agreement with the simulated \hi-rich profile at $R \gtrsim R_1$. The lack of extremely \hi-massive galaxies in \eagle{} is therefore directly connected to the missing \hi{} in the central galaxy regions.  

\subsubsection{Why are the simulated profiles different from observations?}
\label{sec:difference.reason}
We are therefore faced with two puzzles: (i) Why does the inclination correction through elliptical bins work so poorly for the outskirts of the simulated galaxies? (ii) Why is the \hi{} density too low in the inner regions of simulated galaxies? 

For the first question, one natural explanation might be that the inclination of the \hi{} disk does not correspond exactly to the ellipticity of the r-band light (e.g.~\citealt{Serra_et_al_2012, Wang_et_al_2013}). We test this in Appendix \ref{sec:app.incli} and there is indeed good evidence that this is the case: observationally, the distribution of $b/a$ in the \bluedisk{} sample is far from flat, with a deficit at both small ($b/a \lesssim 0.3$) and high ($b/a \gtrsim 0.9$) axis ratios, which is in conflict with the simple assumption $\cos\theta = b/a$ because the distribution of $\cos\theta$ should be uniform. Likewise, a direct comparison between the inclination angle and stellar $b/a$ in \eagle{} shows both significant scatter and a systematic offset for galaxies with $b/a \lesssim 0.6$. However, we have repeated the `elliptical bin' analysis on our simulated galaxies, with axis ratio and orientation angle derived directly from the orientation of the \hi{} angular momentum axis instead of fits to the stellar light, and the resulting profiles are almost identical to those shown in the bottom panel of Fig.~\ref{fig:bd.profiles}. 

There must therefore be a genuine difference between the simulated and observed galaxies, and one obvious candidate for this is the artificially increased thickness of the \hi{} disk in \eagle{} (see discussion in Section \ref{sec:morphology}), which may lead the elliptical-bin inclination correction, based on an infinitesimally thin \hi{} disk, to fail. Any line of sight will intersect a thick disk not only at one point, but over a finite interval, which effectively smears out the resulting profile: an inclined line of sight that intersects the disk mid-plane in the galaxy outskirts (at a radius $R_a$) will actually pick up most \hi{} at $R < R_a$ where the \hi{} density is higher, thus leading to a shallower outer profile. By conservation of total mass, the density must appear lower in the inner regions, exactly as seen in Fig.~\ref{fig:bd.profiles}. This interpretation would also explain why the (outer) face-on profiles in \eagle{} are still a good match to observations, because they are by definition insensitive to the vertical structure of the \hi{} disks. 

It is conceivable, of course, that a similar effect is also present in observed galaxies. However, we have tested this by comparing the (inclination-corrected) profiles for \bluedisk{} galaxies with axis ratio $b/a$ below and above the median of 0.6. If the inclination correction for observed galaxies were to suffer from the same bias as in \eagle{}, the more edge-on sample with $b/a < 0.6$ should show a systematically shallower outer profile, but we did not find evidence for such an offset. The profiles presented by \citet{Wang_et_al_2014} can therefore be regarded as truly face-on, and the comparison to the face-on profiles from \eagle in the top panel of Fig.~\ref{fig:bd.profiles} as the most meaningful test of the simulated \hi{} surface density profiles. 

\begin{figure*}
  \centering
  
   \begin{subfigure}[t]{0.49\linewidth}
    \includegraphics[width=\linewidth]{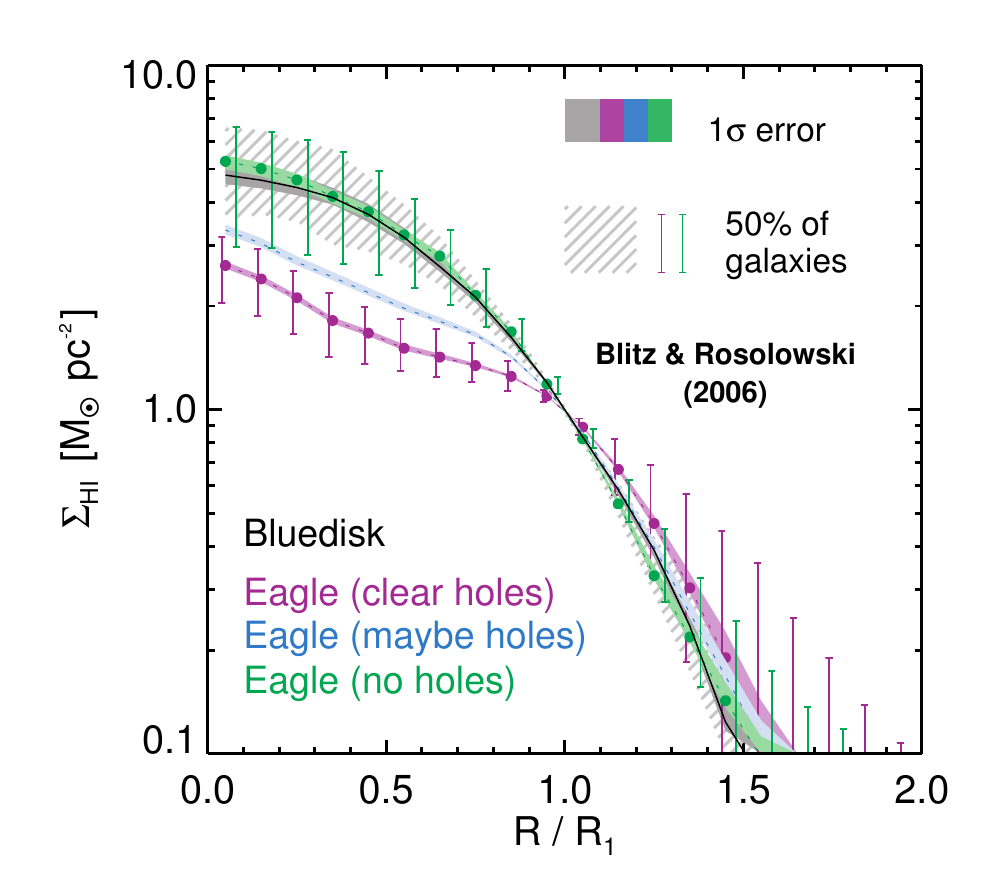}
  \end{subfigure}
  ~
  \begin{subfigure}[t]{0.49\linewidth}
  \includegraphics[width=\linewidth]{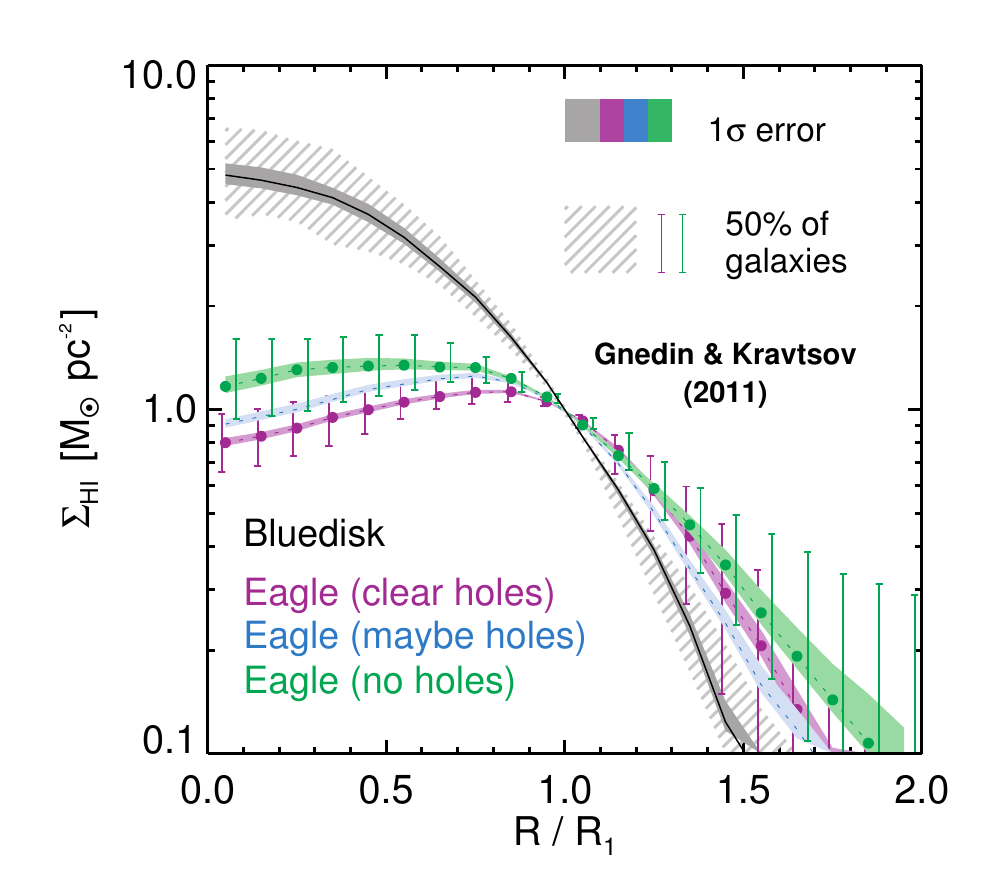}
  \end{subfigure}
  
  \caption{Dependence of \hi{} surface density profiles on the presence of visible \hi{} holes. The \textbf{left-hand} panel shows results obtained with the empirical \citet{Blitz_Rosolowski_2006} H$_2$ model, whereas the theoretical \citet{Gnedin_Kravtsov_2011} formula is used in the \textbf{right-hand} panel. Green points show the profile for galaxies with no visual evidence of holes, while purple points show those galaxies where holes are clearly visible. Galaxies with tentative hole detections are shown in blue; for clarity we have omitted the error bars here. With the empirical H$_2$ formula (left), the hole-free sample (green) is in good agreement with the Bluedisk data over the entire radial range we probe here, while galaxies with hole detections show a deficit in the central \hi{} profile. Using the theoretical formula from \citet{Gnedin_Kravtsov_2011}, galaxies with and without holes have central $\Sigma_\hi$ profiles that are significantly too low compared to observations.}
   \label{fig:bd.profiles_bubbles}
\end{figure*}

The discrepancy in the \emph{central} regions must have a different origin, as it is present in both the face-on and elliptical-bin-corrected profiles. One possibility here is that this is related to the spurious \hi{} holes that we had already noted in the discussion of Fig.~\ref{fig:bd.visual_images}. To test this hypothesis, we show in Fig.~\ref{fig:bd.profiles_bubbles} the median-stacked profiles (now again normalised by $R_1$) for galaxies with clear visual hole detections (purple), as well as those without (green) and those with tentative-only identifications (blue). For simplicity we here combine \hi-rich and control galaxies. The left-hand panel shows the profiles obtained with the empirical BR06 pressure-law to account for the presence of H$_2$; for comparison we also show the equivalent profiles obtained when using the theoretical GK11 formula in the right-hand panel. In both cases, the observed profile from \bluedisk{} (\hi-rich and control combined) is shown in black.

It is evident that the discrepancy in the inner region seen with the BR06 model as discussed above is indeed connected to the presence of \hi{} holes. Simulated galaxies without holes (green) follow the observed profile almost exactly over the entire radial range that we consider here (discrepancy $<$ 0.07 dex). The scatter remains slightly larger than in the observations, but only by typically $\lesssim 50$ per cent. In contrast, galaxies with clearly visible holes (purple) have a much shallower central profile (by up to a factor of $\sim$2), with tentative identifications (blue) lying in-between. It is worth pointing out that the `hole' and `no-hole' populations have profiles that differ by more than the typical scatter in each: there is a clear difference between \emph{individual} galaxies in the two categories, and not just in a statistical sense. 

Although smaller, there is also a slight effect in the outer profiles where the \hi{} surface density is slightly higher in galaxies with holes than without (by $\lesssim 0.1$ dex). This may seem counter-intuitive, but is explained by the fact that the profiles are scaled by $R_1$ which is affected by the presence of holes as well (see Fig.~\ref{fig:bd_r1}). It is, of course, no surprise that galaxies with holes lack \hi{} and therefore show shallower inner profiles than those without visible evidence for them. What is far less trivial, however, is that the `hole-free' sample agrees so well with the observations: \emph{the discrepancy in the surface density profiles between \eagle{} and \emph{\bluedisk{}} can thus be fully attributed to the existence of these holes.}

The GK11 profiles (right-hand panel of Fig.~\ref{fig:bd.profiles_bubbles}) show a similar general trend -- a higher central \hi{} surface density for hole-free galaxies -- but the difference is much smaller here than with the BR06 formula. Moreover, all three profiles are significantly too low within $R_1$, with a discrepancy by a factor of $\gtrsim 5$ in the very centre. Evidently, the theoretically-based GK11 formula assigns an unphysically large fraction of the neutral gas in the central \eagle{} galaxy regions as H$_2$, which would explain why the GK11 total \hi{} masses as shown in Fig.~\ref{fig:gass.masshist} are biased low. In addition, the \emph{outer} $\Sigma_\hi$ profiles are actually slightly more discrepant (i.e.~shallower) for hole-free galaxies than those with holes, indicating that the normalisation radius $R_1$ is somewhat too small at fixed $\mstar$ (see also Fig.~\ref{fig:bd_r1}). The lack of \hi{} in the central region also explains the smaller effect of holes, compared to BR06: there is simply not enough \hi{} even in hole-free galaxies for these features to have a significant impact.

We note that it is, in principle, conceivable that the profile agreement between \hi-rich and control galaxies is not actually a physical feature, but rather a result of the comparatively large beam size: recall that the \hi{} maps from \eagle{} had been artificially reduced in resolution to the same level as in \bluedisk{}. However, we explicitly check for this in Appendix \ref{sec:app.beamsize}, where we calculate the \eagle{} \hi{} surface density profiles from higher-resolution maps. Although the detailed shape of the profiles does indeed vary with the beam size, the close agreement between \hi-rich and control galaxies remains. This strongly suggests that it is a genuine physical feature of the simulated -- and observed -- galaxies, rather than a smoothing artefact. 

The conclusion that the discrepancy between \eagle{} and the Bluedisk observations can be attributed to feedback-related \hi{} holes, which are generated \emph{after} accretion of the gas, combined with the good match to the observed outer \hi{} profile, supports the hypothesis that the accretion of \hi{} onto galaxies is overall modelled well in \eagle. At the same time, it highlights the fact that gas properties of galaxies are particularly sensitive to the adopted feedback implementation in hydrodynamical simulations (see also \citealt{Crain_et_al_2015}).


\section{Summary and Discussion}
\label{sec:summary}

We have conducted a series of detailed like-with-like comparisons between the atomic hydrogen (\hi) content of present-day galaxies with $\mstar \geq 10^{10} \msun$ from the 100 cMpc \eagle{} simulation and various observational data sets. Our main aim has been to test the \eagle{} galaxy formation model in a regime that was not considered during calibration, and to assess the usefulness of these simulations to better understand galactic \hi{} evolution and the origin of observed scaling relations. Our main findings may be summarised as follows:

\begin{enumerate}

\item The \eagle{} simulations, combined with the \citet{Rahmati_et_al_2013} fitting formula for self-shielding and collisional ionisation, predict median neutral hydrogen fractions for central galaxies that agree with observations to better than 0.1 dex at galaxy stellar masses in the range $10^{10}$ to $10^{11}$ $\msun$. For the most massive galaxies ($\mstar > 10^{11} \msun$) large observational uncertainties due to frequent non-detections prevent strong statements on the accuracy of \eagle.

\item Accounting for molecular hydrogen (H$_2$) in \eagle{} with the empirical \citet{Blitz_Rosolowski_2006} pressure law leads to atomic hydrogen masses that are generally in good agreement with observations from \gass; the medians differ by $< 0.2$ dex over a decade in stellar mass from $10^{10}$ to $10^{11} \msun$. A minor deficiency is the failure of the simulation to produce galaxies as \hi-rich as the richest observed galaxies, as well as a moderate shortfall of very \hi-poor galaxies.

\item An alternative, theoretically-based $H_2$ model of \citet{Gnedin_Kravtsov_2011} applied to \eagle{} results in galaxies having overall too little \hi{}, particularly in their centres.

\item Using visual classification of \hi{} morphologies, we have shown that \hi{} disks are increasingly common in simulated galaxies of increasing $\mhi$ and decreasing $\mstar$, but at $\mhi \gtrsim 10^{9.4}\, \msun$ the majority of disks appear vertically disturbed. 

\item Many simulated galaxies contain large (up to $\sim$20 kpc) \hi{} holes which are a factor of several larger than seen in observations. They are more common at high $\mhi$ and low $\mstar$, but show no clear correlation with the specific star formation rate of galaxies.

\item Simulated galaxies match the observed \hi{} mass--size relation reasonably well (the slope is too steep by ~13 per cent), and the agreement becomes excellent (better than 0.1 dex) when only galaxies with visually confirmed \hi{} disks without holes are included and the empirical \citet{Blitz_Rosolowski_2006} prescription is used to account for the presence of H$_2$.

\item The \hi{} surface density profiles of \hi-rich ($\mhi > 10^{9.8} \msun$) and control galaxies ($10^{9.1} \msun \leq \mhi \leq 10^{9.8} \msun$) in \eagle{} follow each other closely when scaled by the characteristic \hi{} radius $R_1$, as observed. While the outer profiles ($R > R_1$) also agree quantitatively, the surface density around $0.5 R_1$ is too low by a factor of $\sim$2. This tension can be fully attributed to the presence of \hi{} holes. Galaxies without holes follow the observed density profile almost perfectly, to better than 0.07 dex.

\end{enumerate}

To our knowledge, this is the first time that such relatively detailed agreement of \hi{} properties with observations has been demonstrated in self-consistent cosmological hydrodynamical simulations. Our results bode well for theoretical studies aiming to use \eagle{} to obtain a better understanding of a wide variety of physical processes relevant for galaxy evolution, such as \hi{} accretion and recycling, as well as its stripping in the dense environment of groups and clusters, for which realistic initial conditions are a significant advantage. In companion papers we show that this success can be predominantly attributed to the calibrated model for energetic feedback from star formation in \eagle{} (\citealt{DallaVecchia_Schaye_2012,Schaye_et_al_2015,Crain_et_al_2015}).

The discrepancy between the empirical and theoretical \hi/H$_2$ partition schemes that we have identified (see also Appendix \ref{sec:app.h2}) is the largest systematic uncertainty in our results. Although the pressure law parameterisation of \citet{Blitz_Rosolowski_2006} is based directly on observational data, its simplicity may hide detailed problems of the \eagle{} simulations that have a strong impact on the other partition models. One possibility here is that the metallicities are somewhat too high in the central galaxy region, which would lead to overestimation of the H$_2$ content by the \citet{Gnedin_Kravtsov_2011} scheme. 

An unambiguous shortcoming of the \eagle{} galaxies in terms of their gas distribution appears to be the common occurrence of unphysically large \hi{} holes with low surface density. Although they appear to be seeded by heating events which are included in \eagle{} to model star formation feedback, the fact that holes are (slightly) more common in galaxies with \emph{low} star formation activity suggests that their formation and survival depends on other galaxy properties as well. Efforts to improve the simulations in this respect must therefore likely involve improvements to both the star formation feedback scheme and ISM model. Another important area for improvement would be to account for the (as yet uncertain) effect of local stellar radiation on the neutral hydrogen fractions: \citet{Rahmati_et_al_2013b} showed that this can significantly affect \hi{} column densities even at $z=0$.

In a follow-up paper, we will study the \hi{} accretion onto \hi-rich \eagle{} galaxies directly by tracing galaxy progenitors back in time to find out when and how the $z = 0$ \hi{} reservoirs were built up. As well as providing new insight into the details of galaxy formation, this will also lead to new predictions that can be tested against future observations, and contribute to our theoretical understanding of galaxies.

\section*{Acknowledgments}
RAC is a Royal Society University Research Fellow. We thank the THINGS, Bluedisk, \gass{} and \coldgass{} teams for making their observations publicly available, Mei-Ling Huang for technical help with the data, and Barbara Catinella for providing the match of \gass galaxies to the \citet{Yang_et_al_2012} halo catalogue. We gratefully acknowledge helpful discussions with Ali Rahmati, Diederik Kruijssen, Thorsten Naab, Bernhard R\"ottgers, Jing Wang, and Simon White. This work used the DiRAC Data Centric system at Durham University, operated by the Institute for Computational Cosmology on behalf of the STFC DiRAC HPC Facility (www.dirac.ac.uk). This equipment was funded by BIS National E-infrastructure capital grant ST/K00042X/1, STFC capital grant ST/H008519/1, and STFC DiRAC Operations grant ST/K003267/1 and Durham University. DiRAC is part of the National E-Infrastructure. We also gratefully acknowledge PRACE for awarding the EAGLE project with access to the Curie facility based in France at Tr\`{e}s Grand Centre de Calcul. Support was also received via the Interuniversity Attraction Poles Programme initiated by the Belgian Science Policy Office ([AP P7/08 CHARM]), the National Science Foundation under Grant No. NSF PHY11-25915, and the UK Science and Technology Facilities Council (grant numbers ST/F001166/1 and ST/I000976/1) via rolling and consolidated grants awarded to the ICC, and the European Research Council (grant agreement 278594 -- GasAroundGalaxies). Several figures in this paper were produced using the \textsc{Astropy} \citep{Astropy_2013} and \textsc{Matplotlib} \citep{Hunter_2007} Python packages.
\bibliographystyle{mn2e}
\bibliography{bibliography_eagle_hi}


\begin{appendix}


\section{Sensitivity to modelling assumptions}
\label{sec:app.methcomp}

\subsection{\hi{} modelling}
\label{sec:app.h2}
We here explore in more detail the impact of adopting alternative prescriptions to model \hi{} in post-processing. Four groups of variations are considered in turn: (i) a different parameterisation of the H$_2$ pressure law; (ii) theoretically-motivated \hi/H$_2$ partition schemes; (iii) more simplistic partition schemes; and (iv) a simple temperature--density cut to identify `\hi' gas particles.

\begin{figure*}
  \centering
  \begin{subfigure}[b]{2\columnwidth}
  \includegraphics[width=\linewidth]{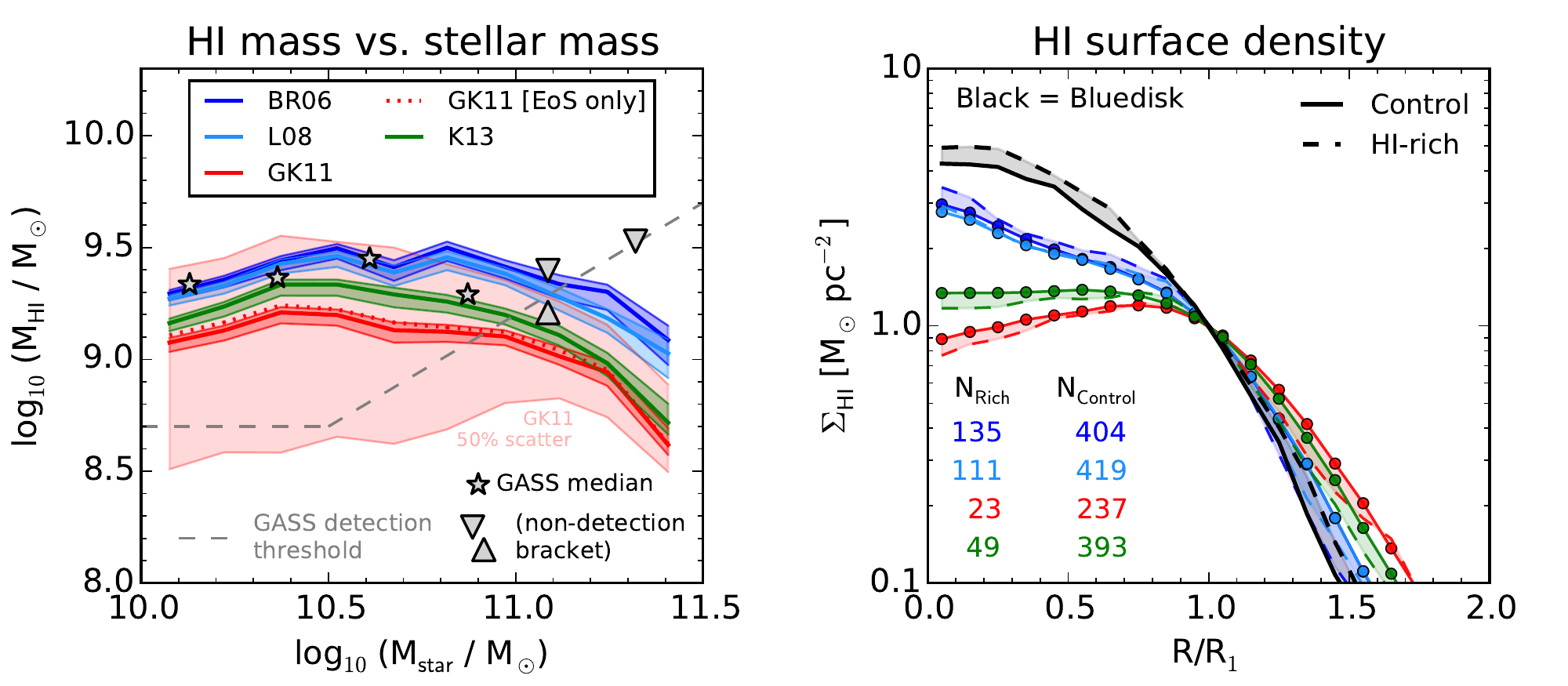}
  \end{subfigure}
  
  \begin{subfigure}[b]{2\columnwidth}
  \includegraphics[width=1\linewidth]{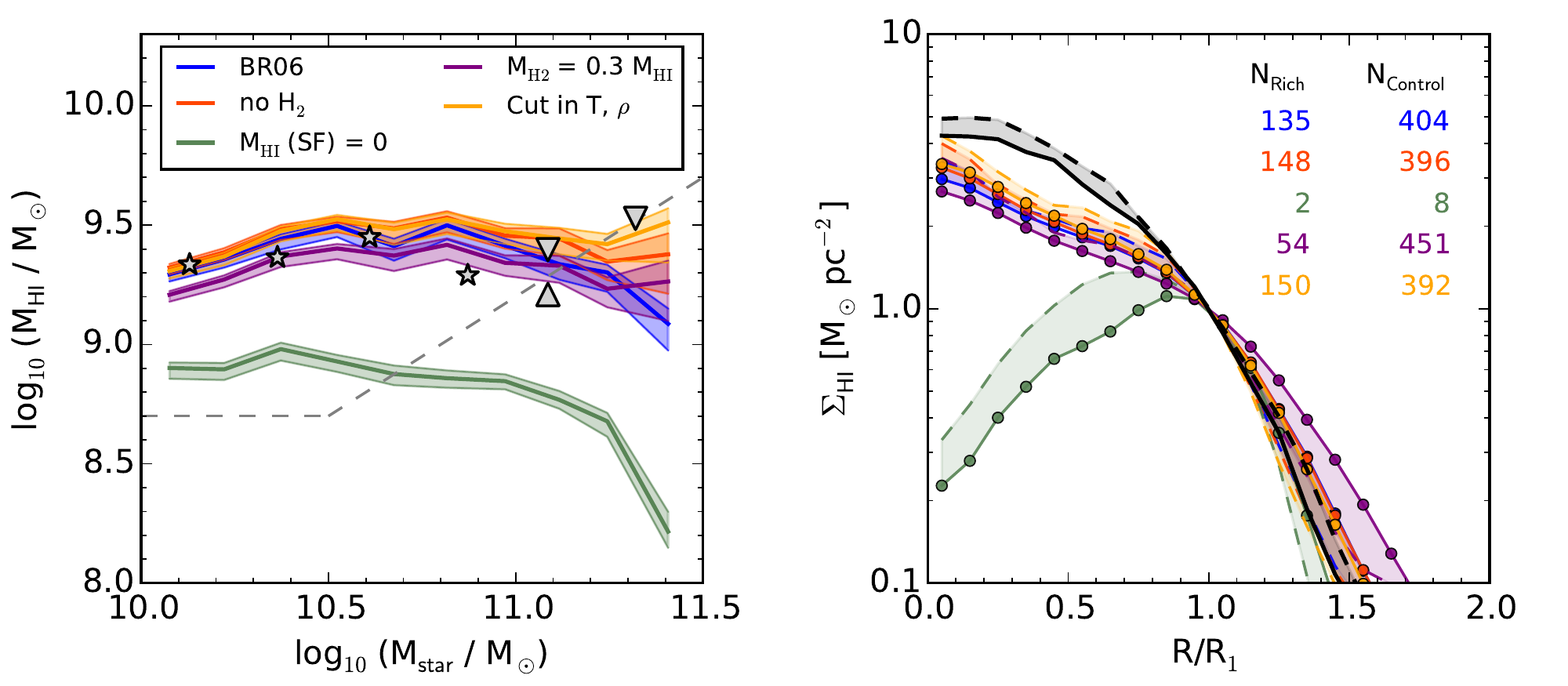}
  \end{subfigure}

       \caption{Comparison of the $\mhi$--$\mstar$ relation (\textbf{left}) and \hi{} surface density profiles (\textbf{right}) predicted by different \hi{} models. \textbf{Top} row: empirical pressure law fit of \citet[light blue]{Leroy_et_al_2008} as well as the theoretically motivated formulae of \citet[red]{Gnedin_Kravtsov_2011} and \citet[green]{Krumholz_2013}. The  \textbf{bottom} row shows more simplistic models: ignoring the presence of H$_2$ (orange), assuming all star forming gas particles are exclusively H$_2$ (green), a fixed ratio of $M_{H_2} = 0.3 \mhi$ as used by \citet[purple]{Popping_et_al_2009}, and a temperature--density cut (yellow). For ease of comparison, the empirical \citet{Blitz_Rosolowski_2006} pressure law model is shown in both panels (dark blue). Lines show median values, the dark shaded bands in the right panels indicate their 1$\sigma$ uncertainty. For the \citet{Gnedin_Kravtsov_2011} model in the top (red), we additionally show the 50 per cent scatter with a light-shaded band as explained in the text; this is similar for the other models and hence not shown for clarity. Numbers in the right-hand panels show how many \hi-rich and control galaxies are predicted by each model.}
    \label{fig:app.h2}
  \end{figure*}

For each of these, we show in Fig.~\ref{fig:app.h2} the resulting $\mstar$--$\mhi$ relation (left-hand panels) and scaled \hi{} surface density profiles in analogy to Fig.~\ref{fig:bd.profiles} (right-hand panels). Solid lines represent running medians; the \citet{Blitz_Rosolowski_2006} pressure law is shown in dark blue in both rows for ease of comparison. 

In the right-hand plots, the dark shaded bands represent the 1$\sigma$ uncertainty on the median (see Section \ref{sec:profiles.comparison}). The 50 per cent galaxy-to-galaxy scatter is shown for the \citet{Gnedin_Kravtsov_2011} model (GK11) in the top row as a light red band; for the other models this scatter is similar and hence not shown for clarity. Observed median \hi{} masses from \gass{} \citep{Catinella_et_al_2013} are over-plotted as grey symbols. In analogy to Fig.~\ref{fig:neutralgas}, we consider setting non-detections in \gass{} both to zero and to the observational upper limit. Where the median is the same in both approaches (i.e. where its value is unaffected by the presence of non-detections), we show it as a star-symbol, otherwise the two differing values which bracket the true median are shown by downward and upward facing triangles\footnote{The reason for being able to compute a non-zero lower limit in the second-highest stellar mass bin which differs from the upper limit is that 52 per cent of these galaxies are detected in \hi{} by \gass{}, but some non-detected galaxies have upper limits higher than the lowest detected \hi{} masses; see \citet{Catinella_et_al_2010} for details. In the highest stellar mass bin, only 36 per cent of galaxies are detected so the lower limit on the median is zero.}. The grey dashed line shows the nominal \gass{} detection threshold.

In the right-hand plots, ``control'' (log$_{10}$ ($\mhi/\msun$) = [9.1, 9.8]) and ``\hi-rich'' galaxies (log$_{10}$ ($\mhi/\msun$) $\geq 9.8$) are differentiated by solid and dashed lines, respectively; their numbers are shown in the bottom-left or top-right corners. For comparison, the observational data from \bluedisk{} \citep{Wang_et_al_2014} are shown in black. Shaded bands highlight the extent of discrepancy between the \hi-rich and control profiles.

\subsubsection{Different parameterisations of the H$_2$ pressure law}
The light blue lines in the top panels represent the results obtained with the alternative pressure law parameterisation of \citet[L08]{Leroy_et_al_2008}. Their analysis is based on the \emph{THINGS} survey \citep{Walter_et_al_2008} with a larger sample of galaxies (23) than that of \citet[11 galaxies]{Blitz_Rosolowski_2006}. The best-fit parameters determined by L08 are both slightly lower than in BR06: $P_0/k_B = 1.7\times 10^4$ cm$^{-3}$ K and $\alpha = 0.8$ (BR06: $P_0/k_B = 4.3\times10^4$ cm$^{-3}$ K and $\alpha = 0.92$). However, the impact on our results is negligible, and both methods agree to better than 0.05 dex except at the high-$\mstar$ end (where the molecular fraction is highest; \citealt{Saintonge_et_al_2011}). Likewise, the central $\Sigma_\hi$ profiles are lower in L08 (especially for \hi-rich galaxies), but only by $\lesssim 10$ per cent. We therefore conclude that our results are not significantly affected by uncertainties in the H$_2$ pressure law parameterisation.

\subsubsection{Theoretically motivated \hi/H$_2$ partition schemes}
In addition to the \citet[``GK11'']{Gnedin_Kravtsov_2011} formula, \citet{Lagos_et_al_2015} also implemented the theoretically motivated \hi/H$_2$ partition schemes of \citet[``K13'']{Krumholz_2013} into \eagle, which is derived from analytic modelling; as with GK11, we refer to \citet{Lagos_et_al_2015} for details of the implementation. In the top row of Fig.~\ref{fig:app.h2}, the K13 model is shown in dark green, and the corresponding result from GK11 in red.

In contrast to the L08 parameterisation discussed above, the K13 model predicts \hi{} masses that are too low compared to observations (by $\sim$0.2 dex at $\mstar \approx 10^{10} \msun$ and $\sim$0.5 dex at the high-$\mstar$ end). They are also lower than the predictions from either of the two pressure law parameterisations (see above), but consistently higher that what is predicted by the GK11 formula (red, a difference of $\sim$0.1--0.2 dex). As the top-right panel of Fig.~\ref{fig:app.h2} shows, this discrepancy is predominantly driven by the inner galaxy regions ($R \lesssim R_1$) where K13 (as well as GK11) combined with \eagle{} fails to predict (median) surface densities above 1.5 $\msun$ pc$^{-2}$ (lower by a factor $\sim$4 than observed by \bluedisk).

We also note that the \citet{Lagos_et_al_2015} implementation of GK11 and K13 allows for the existence of a (small) molecular fraction in all gas particles, including those not forming stars. In contrast, the pressure law models of BR06 and L08 are only applied to star forming particles whose properties admit the formation of a significant amount of cold gas \citep{Schaye_2004, Schaye_DallaVecchia_2008, Schaye_et_al_2015}. To demonstrate that this is of little significance for the resulting \hi{} content ($< 0.05$ dex), the median $\mhi$ obtained with the GK11 prescription but H$_2$ restricted to star forming particles is shown in the top-left panel as a red dotted line.

\subsubsection{Simple \hi/H$_2$ partition schemes}
The \hi/H$_2$ partition schemes considered so far all have a particle-level physical basis, either on theoretical or observational grounds. Three more simple separation prescriptions are considered in the bottom panel of Fig.~\ref{fig:app.h2}: ignoring the presence of H$_2$ altogether (red), assuming a fixed ratio of $M_{\text{H}_2}/\mhi = 0.3$ (purple), which approximately corresponds to the cosmic average at $z = 0$ for the galaxy masses considered here \citep{Saintonge_et_al_2011} and was used by \citet{Popping_et_al_2009}; and an extreme prescription of assuming \hi{} is only found in non star forming particles (which, in turn, contain no H$_2$; pale green).

As expected, the latter prescription strongly under-predicts (by $\sim$0.5--1 dex) the \hi{} content of simulated galaxies, even more than the GK11 or K13 partition schemes and shows an unphysical, strong central decline in $\Sigma_\hi$ (because most gas inside $R_1$ has a non-zero star formation rate in the simulation). In contrast, the other two agree relatively closely with BR06 (dark blue), deviating by less than 0.1 dex in $\mhi$ except for the few galaxies with $\mstar > 10^{11.2} \msun$. By construction, the model ignoring H$_2$ predicts slightly higher $\mhi$ whereas the fixed particle-level $M_{\text{H}_2}/\mhi$ ratio (purple) generally yields slightly lower integrated $\mhi$ except for the comparatively H$_2$ rich most massive galaxies. Both are compatible with the observational constraints from \gass. 

These trends are mirrored by the $\Sigma_\hi$ profiles, except that the profiles for `control' galaxies are slightly too shallow with fixed particle-level $M_{\text{H}_2}/\mhi$ ratio (purple), plausibly because this assumption breaks down in the (molecule-poor) outer parts and therefore underestimates the true $R_1$ radius.

\subsubsection{Temperature--density cut for \hi}
Finally, we test the arguably simplest model of assuming a cut in temperature\footnote{We use a fixed value of $T = 10^4$ K for star-forming gas here.} ($T \leq 10^{4.5}$ K) and density (n$_\text{H} \geq $ 0.01 cm$^{-3}$); we then set $m_\hi = m_\text{H}$ for these particles and $m_\hi = 0$ for all others. In contrast to all other models explored here, this prescription does not rely on the \citet{Rahmati_et_al_2013} fitting formula for calculating the neutral hydrogen fraction, and is shown in yellow in the bottom row of Fig.~\ref{fig:app.h2}. Although both the integrated $\mhi$ and surface density profiles are slightly higher than with the empirical BR06 approach (dark blue), this difference is similarly small as for the ``ignoring H$_2$'' model (orange) and deviates by less than 0.1 dex from our default prescription except for the most massive galaxies.

\subsection{Integration aperture}
\label{sec:app.aperture}
\begin{figure}
  \centering
    \includegraphics[width=\columnwidth]{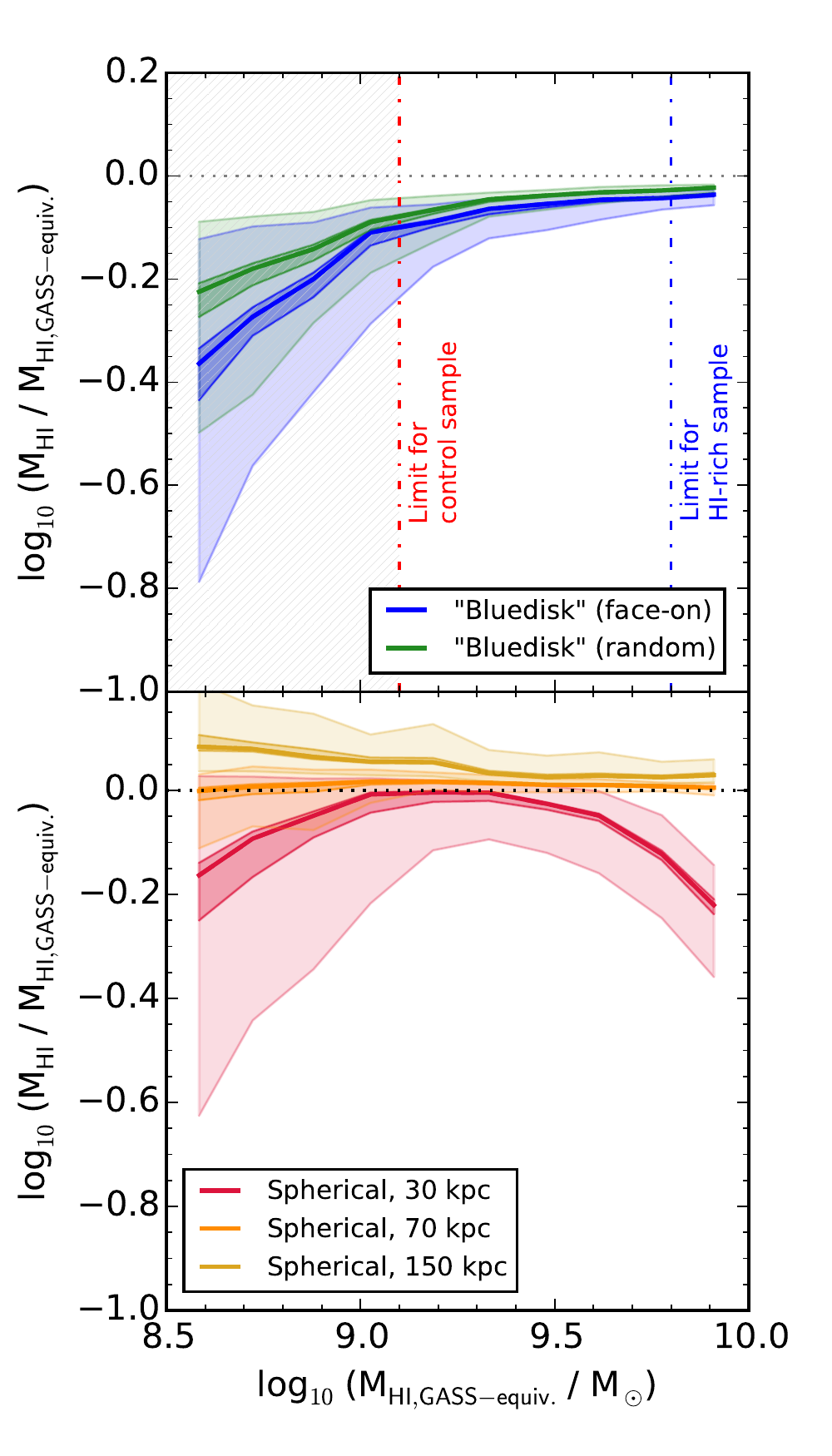}
       \caption{Difference between the ``\gass-equivalent'' \hi{} measurement (within a 2D aperture of 70 kpc, and $|\Delta v| \leq 400$ km s$^{-1}$), and the mass obtained from \hi{} images at the \bluedisk{} resolution with a threshold \hi{} surface density of $\Sigma_\text{thresh} = 0.37\, \msun\, \text{pc}^{-2}$ (top panel). Thin shaded bands enclose 50\% of galaxies in random orientation (green) or face-on (blue), while the darker shaded bands indicate the statistical $1\sigma$ uncertainty on the median (solid line). The difference is less than 10 per cent for galaxies with \hi-masses above the `control' limit ($\mhi \geq 10^{9.1} \msun$; red line). For reference, the light blue vertical line indicates the division between \hi-rich and control galaxies in Section \ref{sec:profiles}. Bottom: Comparison to \hi{} masses in spherical apertures of 30 kpc (red), 70 kpc (orange), and 150 kpc (yellow).}
    \label{fig:app.mhicomparison}
  \end{figure}

In Fig.~\ref{fig:app.mhicomparison}, we test to what extent the \hi{} masses obtained using our default ``\gass-equivalent'' mass definition (R$_\text{2D} \leq 70$ kpc and \( \lvert\Delta v_z \rvert \leq 400 \) km s$^{-1}$) agree with other choices. The top panel compares to the ``\bluedisk-equivalent'' definition (see Section \ref{sec:bd.profiles}):  
we make a $140\times140$ kpc mock \hi{} image from all particles with $\lvert z \rvert \leq 70$ kpc, convolve this with a Gaussian $14 \times 9$ kpc FWHM beam, and then integrate over all pixels with $\Sigma_\hi \geq 0.37 \msun$ pc$^{-2}$ (as in \citealt{Wang_et_al_2013}). Blue and green lines show the median \hi{} mass difference for galaxies in face-on and random orientation, respectively (dark and light shaded bands give the $1\sigma$ uncertainty on the median and 50 per cent scatter). Although the ``\bluedisk'' method gives somewhat lower \hi{} masses, the difference is only $\sim$0.1 dex (25 per cent) at the lower limit of our `control' sample ($\mhi = 10^{9.1}\,\msun$, vertical red dash-dot line), and less than 0.05 dex (12 per cent) for ``\hi-rich'' galaxies ($\mhi \geq 10^{9.8} \msun$). Our results presented above are therefore not significantly affected by the difference between these two integration methods (see also Fig.~\ref{fig:bd_r1}).

In the bottom panel, we compare to simple 3D radial cuts at 30 kpc (red), 70 kpc (orange), and 150 kpc (yellow). The second of these agrees very well with our ``\gass-equivalent'' aperture down to \hi{} masses as low as $10^{8.5}\,\msun$: only 13 per cent of all galaxies deviate by more than 25 per cent. At intermediate \hi{} masses ($\mhi \approx 10^{9.5} \msun$) the spherical cut leads to a slightly \emph{higher} \hi{} mass: this is because some particles within the sphere have peculiar $z$-velocities larger than 400 km s$^{-1}$ and are therefore excluded in the ``\gass-equivalent'' measurement (the excess is only of the order of a few per cent, however). Adopting a smaller aperture (30 kpc, red) leads to much more significant mass deficits of $\sim$0.2 dex (60 per cent) at both the high- and low-$\mhi$ ends. Perhaps surprisingly, the masses still agree well at intermediate $\mhi$. This behaviour is likely due to an interplay of two effects: at masses of $\mhi \gtrsim 10^{9.5}\, \msun$, $R_1$ is typically 20 kpc or more (Fig.~\ref{fig:bd_r1}), so a 30 kpc aperture misses a non-negligible amount of $\hi$ in the galactic outskirts (see Fig.~\ref{fig:bd.cummass}). For \hi{} poor galaxies ($\mhi \lesssim 10^{9.0}\, \msun$), on the other hand --- which are predominantly of ``irregular'' \hi{} type (Fig.~\ref{fig:bd.morphtypes}) --- a significant contribution to the ``\gass-equivalent'' \hi{} mass comes from \hi{} clumps along the line of sight which are not directly connected to the galaxy. This would also explain why these galaxies show slightly smaller-than-expected $R_1$ radii in Fig.~\ref{fig:bd_r1}: it is really their \hi{} masses that are slightly too high.

Overall, we conclude that our \hi{} masses are not just compatible with the respective observations, but also physically meaningful at least for galaxies with $\mhi > 10^9\, \msun$ upon which the majority of our analysis here is based.


\section{Tests of numerical convergence}
\label{sec:app.rescomp}
In this appendix, we test the numerical convergence of our results, by comparing three additional simulations from the \eagle{} suite run in a 25 cMpc periodic box. Two of these (Ref-L025N0752 and Recal-L025N0752 in the terminology of \citealt{Schaye_et_al_2015}) were run at eight times higher mass resolution (i.e.~$m_\text{gas} \simeq 2.2 \times 10^5 \msun$), while the third one (Ref-L025N0376) uses the same resolution as in the large Ref-L100N1504 simulation that we have analysed in the main part of this paper. The difference between the two high-resolution simulations is that Ref-L025N0752 uses the same sub-grid physics parameters as run Ref-L100N1504 (and Ref-L025N0376), whereas in Recal-L025N0752, the parameters were re-calibrated to improve the match to the observed galaxy stellar mass function. For more details, the interested reader is referred to \citet{Schaye_et_al_2015}.

The rationale behind this re-calibration is that, as explained in detail by \citet{Schaye_et_al_2015}, the interpretation of the numerical sub-grid physics parameters in a hydrodynamical simulation is in general resolution dependent. This makes it unlikely that a simulation such as \eagle{} can achieve ``strong convergence'' -- i.e. the calculation result being unaffected by a change in resolution while keeping the sub-grid parameters fixed. However, one may still obtain ``weak convergence'' by re-calibrating the sub-grid parameters at the new resolution level. 

\begin{figure*}
  \centering
  \includegraphics[width=2.1\columnwidth]{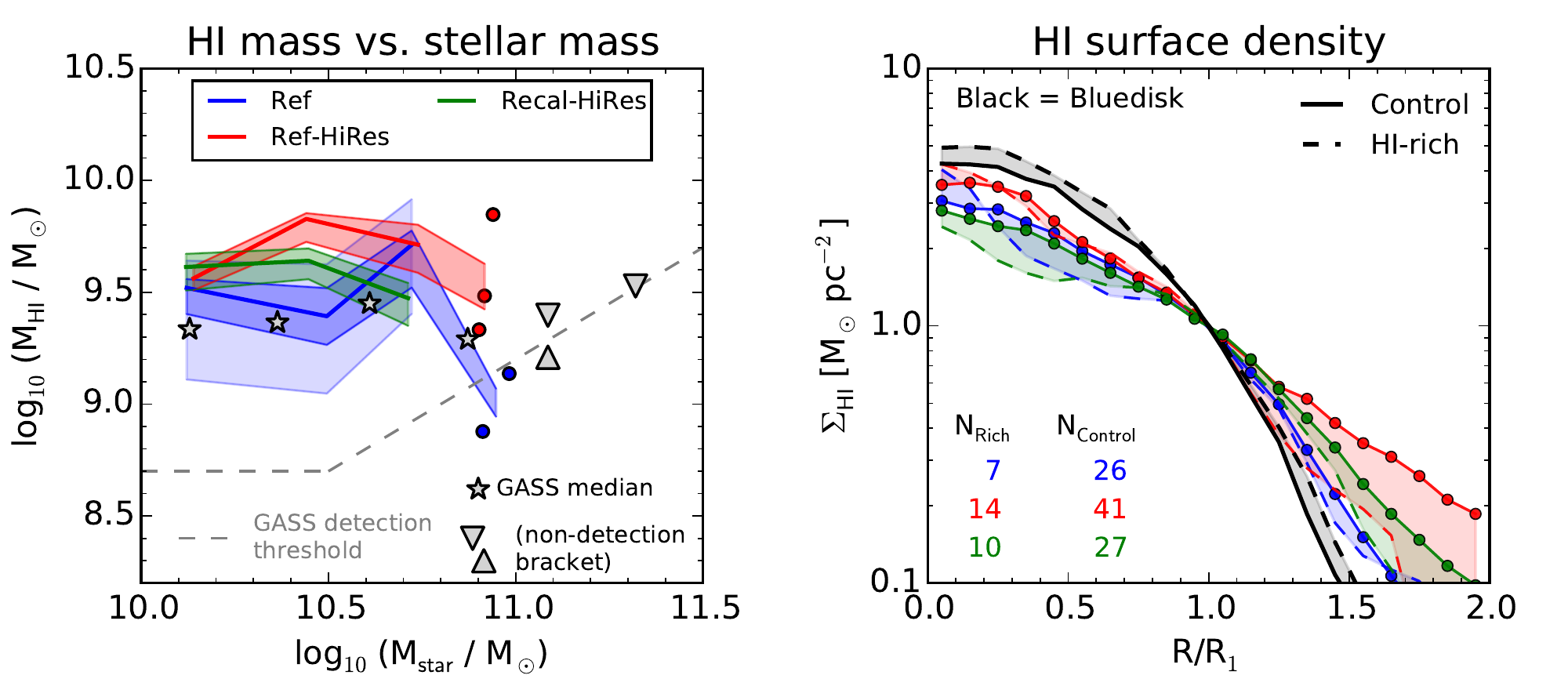}
      
       \caption{Strong and weak convergence tests for \eagle{} \hi{} properties. The layout is analogous to Fig.~\ref{fig:app.h2} above. We compare three 25 cMpc simulations: Ref-L025N0376 (blue; same resolution and parameters as the large run analysed in the main part of the paper), Ref-L025N0752 (red, higher resolution but same parameters), and Recal-L025N0752 (green, higher resolution and re-calibrated parameters). See text for details. In both higher resolution simulations, galaxies are more \hi{} rich, particularly in the galaxy outskirts (shallower profiles in right-hand panel). As expected, this difference is less strong in the re-calibrated simulation (green).}
    \label{fig:rescomp}
  \end{figure*}

We will test both the strong convergence of our results, by comparing the two Ref-L025 simulations (standard and high resolution), and weak convergence by comparing Ref-L025N0376 and Recal-L025N0752. Both these tests are presented in Fig.~\ref{fig:rescomp}, the structure of which is identical to Fig.~\ref{fig:app.h2} above. Note, though, that due to the much smaller volume of the 25 cMpc boxes compared to the large 100 cMpc run (by a factor of 4$^3 = 64$), the number of galaxies is significantly smaller here, resulting in larger statistical uncertainties.

Looking first at the total \hi{} masses (left-hand panel), the general trend is that the high-resolution simulations contain galaxies that have higher $\mhi$ at a given $\mstar$. This difference is larger for simulation Ref-L025N0752 (red), indicating that the weak convergence (blue/green curves; offset by $\lesssim$0.2 dex) is better than strong convergence (blue/red; offset by $\lesssim 0.4$ dex), as expected. However, even the recalibrated high-resolution simulation is significantly too \hi-rich at least at stellar masses between $10^{10.0}$ and $10^{10.5} \msun$ compared to observations.

These trends are mirrored by the \hi{} surface density profiles shown in the right-hand panel of Fig.~\ref{fig:rescomp}. In the central region ($R \lesssim R_1$), agreement between all three simulations is quite good, with the biggest discrepancy again between the two Ref simulations (strong convergence test; high-resolution simulation higher by $\sim$0.2 dex). In the outskirts, both high-resolution simulations show profiles that are shallower than the standard resolution run Ref-L025N0376, indicating extra \hi{} at increased resolution. Again, this difference is stronger for the non-recalibrated high-resolution run Ref-L025N0752 (red), in particular for ``control'' galaxies with $\log_{10}(\mhi/\msun) = [9.1, 9.8]$ where the offset reaches $\sim$0.5 dex. Although the agreement is better for the recalibrated high-resolution run (green), an offset of $\sim$0.2 dex remains at $1.5 R_1$. Furthermore, neither higher resolution simulation reproduces the similarity between the outer \hi{} profiles of \hi-rich and control galaxies seen in \bluedisk{}, although here, as well, the effect is ameliorated by the parameter recalibration: at $1.5 R_1$, the difference between $\Sigma_\hi$ in \hi-rich and control galaxies is $\sim$0.3 dex in Ref-L025N0752, and only $\sim$0.1 dex in Recal-L025N0752.

In summary, both \hi{} masses and surface density profiles in \eagle{} show reasonably good weak convergence, with differences at a level of $\sim$0.1 -- 0.2 dex in the sense that higher-resolution galaxies contain more \hi{}. Strong convergence is considerably less good, with differences up to $\sim$0.5 dex. It is conceivable that further parameter fine-tuning might improve the weak convergence in terms of \hi{} properties, but with typical observational uncertainties at a level of $\sim$0.1 -- 0.2 dex, it is questionable whether this would actually be justified. A more detailed discussion of convergence of \hi{} masses in \eagle{} will be presented by Crain et al. (in prep.).


\section{Bluedisk equivalent sample selection}
\label{sec:app.bdsel}

Unlike \gass{}, the \bluedisk{} sample is by construction biased. The actual selection procedure is somewhat complex (see \citealt{Wang_et_al_2013} for details) and involves a large range of galaxy parameters including stellar mass, stellar surface mass density, NUV and r band colours as well as colour gradients. Although it is, in principle, possible to mimic all these for our simulated galaxies, the risk of many small deviations adding up to major inconsistencies is considerable, especially because of the complex influence of dust on the NUV luminosities. However, we demonstrate here that the selection can be reproduced with a simple size cut of $R_{50,z} = 3\, \text{kpc}$, and a threshold at $\log_{10}\, (M_\textsc{Hi} / \msun) = 9.8$ to separate ``\hi-rich'' and ``control'' galaxies.

\subsection{Overall sample selection by galaxy size}
\label{sec:app.bdsel-size}

\bluedisk{} targeted 25 galaxies expected to be \hi-rich, selected out of those for which the photometric gas fraction plane of \citet{Catinella_et_al_2010} with an additional colour gradient correction \citep{Wang_et_al_2013} predicted an \hi{} mass fraction $\geq 0.6$ dex above the median at a given stellar mass \citep{Catinella_et_al_2010}. The \hi{} mass fraction is strongly anti-correlated with stellar surface mass density $\mu_* \equiv \mstar / (2 R_{50,z}^2)$, so these target galaxies have preferentially low $\mu_*$.\footnote{$R_{50,z}$ is defined as the projected radius enclosing 50 per cent of the Petrosian flux in the z-band. The Petrosian flux, in turn, is the total flux within two Petrosian radii $r_P$. Note that, in SDSS, Petrosian fluxes in all five bands are based on $r_P$ as measured in the r-band, to avoid aperture bias effects.} An equal number of control galaxies were selected to match the \hi-rich targets in $\mu_*$ (amongst other properties), so these are also preferentially of low stellar surface density. 

\begin{figure}
  \centering
    \includegraphics[width=\columnwidth]{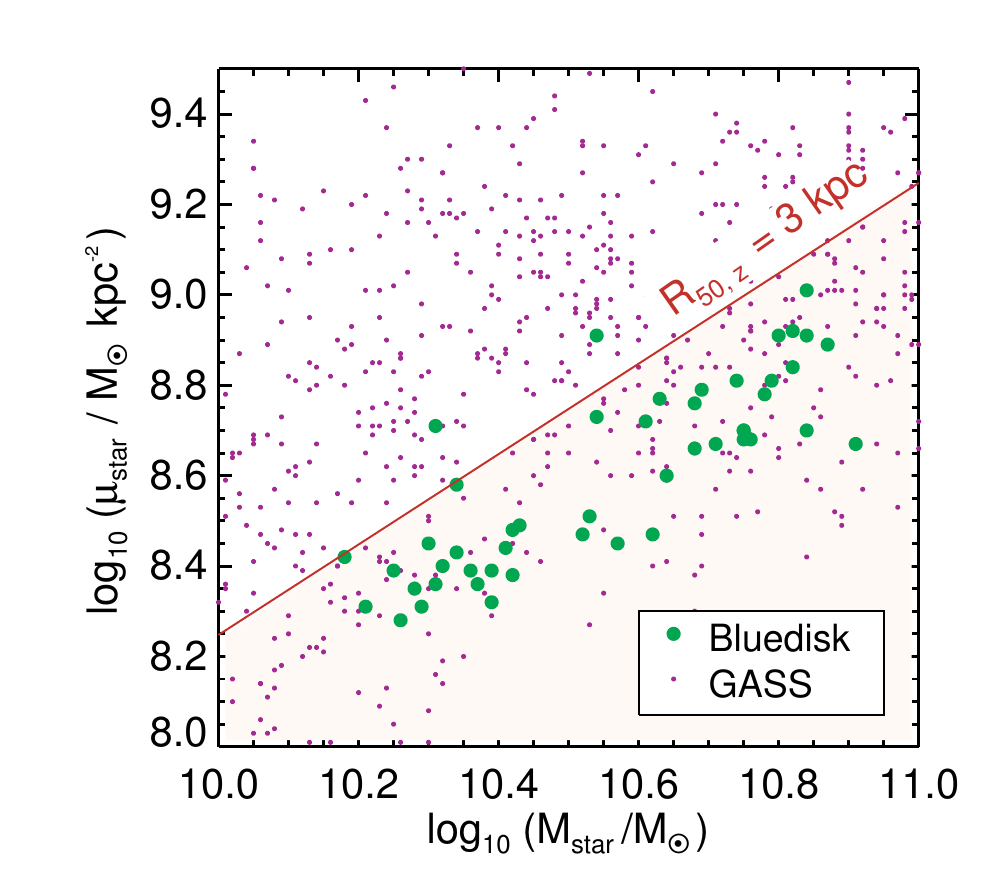}
       \caption{Stellar surface density vs.~stellar mass for \bluedisk{} galaxies (green) and the \gass{} sample (purple). Compared to the unbiased \gass{} sample, \bluedisk{} galaxies have preferentially lower surface densities, i.e.~larger sizes.}
    \label{fig:app.bdsel.size}
  \end{figure}

In Fig.~\ref{fig:app.bdsel.size} we plot $\mu_*$ against stellar mass, $\mstar$, for all \bluedisk{} galaxies (large green circles), and also for the ($\mu_*$-unbiased) \gass{} sample in purple. As expected, the \bluedisk{} galaxies are congregated at the lower end of the \gass{} distribution, with an upper boundary corresponding approximately to a fixed size of $R_{50,z} = 3$ kpc (only two out of 50 galaxies are slightly smaller). 

\begin{figure}
  \centering
    \includegraphics[width=\columnwidth]{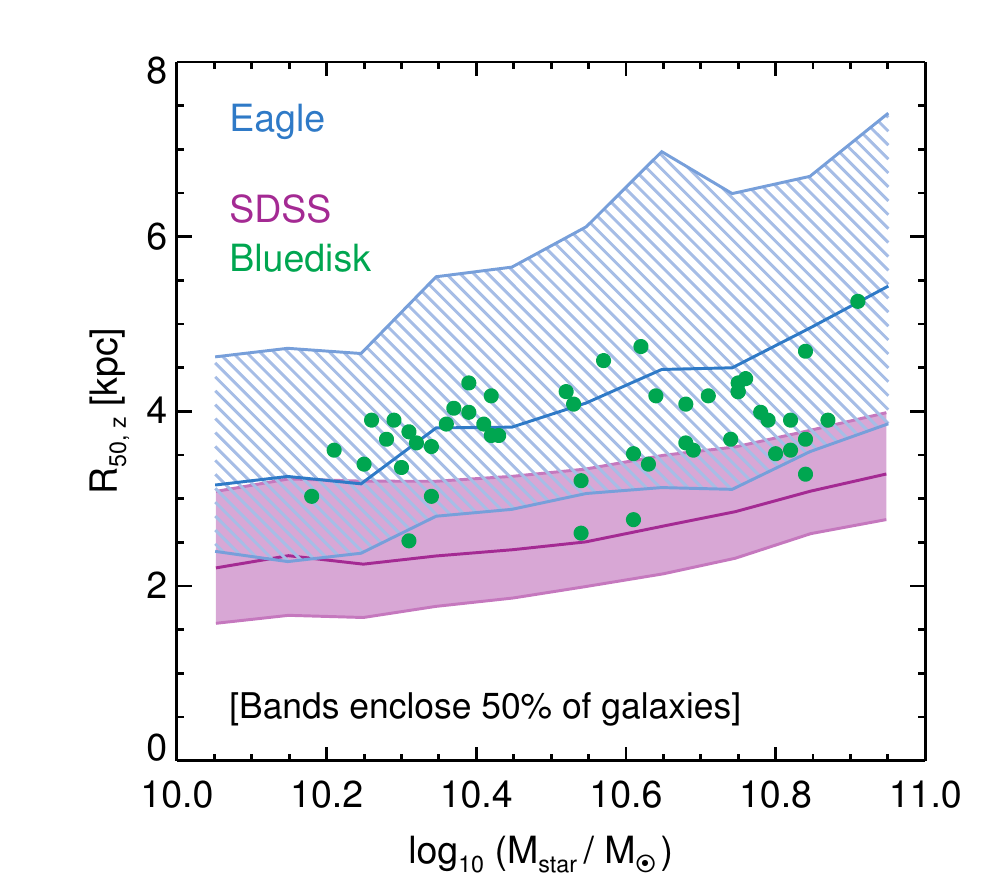}
       \caption{SDSS z-band Petrosian half-light radii $R_{50,z}$ in \eagle{} (blue) and SDSS (\gass{} parent sample; purple). The sizes of the simulated galaxies appear systematically larger than in SDSS, the difference being comparable to the typical size excess of the \bluedisk{} galaxies (green). Note that the size parameter $R_{50,z}$ used here is not the same as that shown by \citet{Schaye_et_al_2015}.}
    \label{fig:app.bdsel.petro}
  \end{figure}

We have therefore calculated the half-light radius $R_{50,z}$ for our simulated galaxies, using luminosities calculated for each star particle using stellar population synthesis \citep{Trayford_et_al_2015}. However, as we show in Fig.~\ref{fig:app.bdsel.petro}, the resulting sizes are systematically larger than in SDSS, of which we take the \gass{} parent sample from \citet{Catinella_et_al_2010} with almost 12,000 galaxies as a large, unbiased subset. The origin of this discrepancy is not entirely clear, because \citet{Schaye_et_al_2015} have demonstrated that the $z=0.1$ stellar sizes in \eagle, when calculated based on S\'{e}rsic profile fits to stellar mass profiles as described by \citet{McCarthy_et_al_2012b}, are in good agreement with the SDSS-based results of \citet{Shen_et_al_2003}. We note here that both the observational size--mass relation of \citet{Shen_et_al_2003} and the \eagle{} relation shown in \citet{Schaye_et_al_2015} are based on galaxies with S\'{e}rsic index $n_S < 2.5$. On the other hand, \citet{Baldry_et_al_2012} find a size--mass relation for blue galaxies that is $\sim$0.2 dex \emph{higher} than the result of \citet{Shen_et_al_2003}, and \citet{Dutton_et_al_2011} have shown that the \citet{Shen_et_al_2003} sizes are biased low due to their use of circular apertures. Our Fig.~\ref{fig:app.bdsel.size}, on the other hand, does not select galaxies by any other criterion than stellar mass. 

It is also possible that the observational analysis in the SDSS pipeline slightly underestimates the true Petrosian radius (e.g.~due to limited depth of the images), that the z-band profiles differ from the stellar mass profiles for \eagle{}, or otherwise that the (outer) stellar light profiles in the simulations are somewhat too shallow. For the present purpose, we simply re-scale the $R_{50,z}$ radii of the \eagle{} galaxies so that its distribution function matches that of the SDSS sample; from Fig.~\ref{fig:app.bdsel.petro} it can be seen that this typically corresponds to a reduction by a factor of $\sim 40$ per cent. The \eagle{} galaxy sizes and their evolution are studied in detail by \citet{Furlong_et_al_2015b}. We emphasize, however, that this discrepancy has virtually no impact on our results here, because we \emph{only} use stellar sizes to select the overall subsample to compare to \bluedisk{} and have verified that our results are virtually unchanged when we instead select all \eagle{} galaxies, regardless of size.

\subsection{Division into \hi-rich and control galaxies by total \hi{} mass}
\label{sec:app.bdsel-rich}

After observation, the \bluedisk{} galaxies were (re-)classified into \hi-rich and control samples based on a comparison between the actual observed \hi{} mass ($M_\text{\hi, observed}$) and that predicted by the original  \citet{Catinella_et_al_2010} gas fraction plane ($M_\text{\hi, C10-prediction}$). Galaxies with \hi{} mass larger than predicted are classified as \hi-rich, while those whose \hi{} mass is less than predicted were assigned to the control sample (compared to all galaxies at a given stellar mass, these are still slightly \hi{} enhanced because of the aforementioned bias towards low stellar surface densities)\footnote{As \citet{Wang_et_al_2013} show, this effect is largely cancelled out by a bias in the \citet{Catinella_et_al_2010} gas fraction plane, which over-predicts the \hi{} content of most galaxies targeted as the `control' sample. Their (actual) $\mhi$ is therefore not far from the overall median.}.

\begin{figure}
  \centering
    \includegraphics[width=\columnwidth]{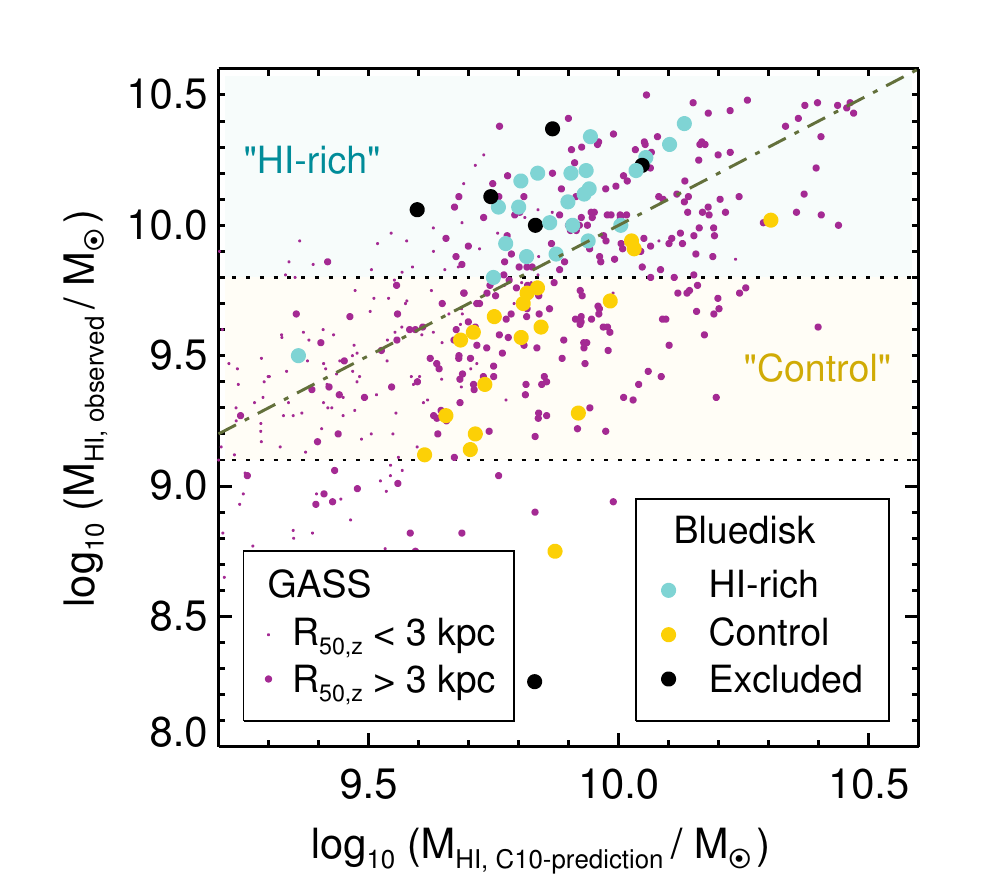}
       \caption{Definitions of `\hi-rich' and `control' galaxies in the \bluedisk{} survey (blue and yellow points) and the simplified sample division adopted here (horizontal dotted lines and shaded regions). The observational division is reproduced almost exactly by these simple cuts.}
    \label{fig:app.bdsel.richsel}
  \end{figure}

In Fig.~\ref{fig:app.bdsel.richsel}, we plot the observed and predicted \hi{} masses of \bluedisk{} galaxies, with \hi-rich galaxies ($M_\text{\hi, observed} > M_\text{\hi, C10-prediction}$) shown in blue and control galaxies ($M_\text{\hi, observed} < M_\text{\hi, C10-prediction}$) in yellow. As can be seen, the two samples can be relatively cleanly separated by a single cut at $M_\text{\hi, observed} = 10^{9.8} \msun$: There are only two \hi-rich galaxies whose \hi{} mass is below this threshold (one of them only marginally), and conversely only three galaxies in the \bluedisk{} control sample lie above the threshold. At the lower end, the \bluedisk{} control sample is well limited by $M_\text{\hi, observed} \geq 10^{9.1} \msun$, with again only two galaxies falling outside this range (one of which is excluded from analysis because of a nearby companion, see \citealt{Wang_et_al_2013}). 

To test our overall sample selection (see above), we also plot the \gass{} galaxies as purple circles, large ones representing `big' \gass{} galaxies ($R_{50,z} \geq 3$ kpc, in the same range as \bluedisk{}), and all others shown as small points. Encouragingly, the first set is also located in the same region as the \bluedisk{} galaxies in this plot, whereas the second set (galaxies smaller than \bluedisk{}) form a tail towards the lower left (i.e.~towards lower \hi{} masses).


\section{Inclination of the \hi{} disk and optical ellipticity}
\label{sec:app.incli}

We had noted in the discussion of Fig.~\ref{fig:bd.profiles} that differences in the outer \hi{} surface density profiles emerge depending on whether the inclination correction is performed using the angular momentum axis of the \hi{}, or the optical r-band ellipticity. We here test the possibility that this is due to incorrect alignment of the \hi{} disks in the latter approach.

\begin{figure}
  \centering
    \includegraphics[width=\columnwidth]{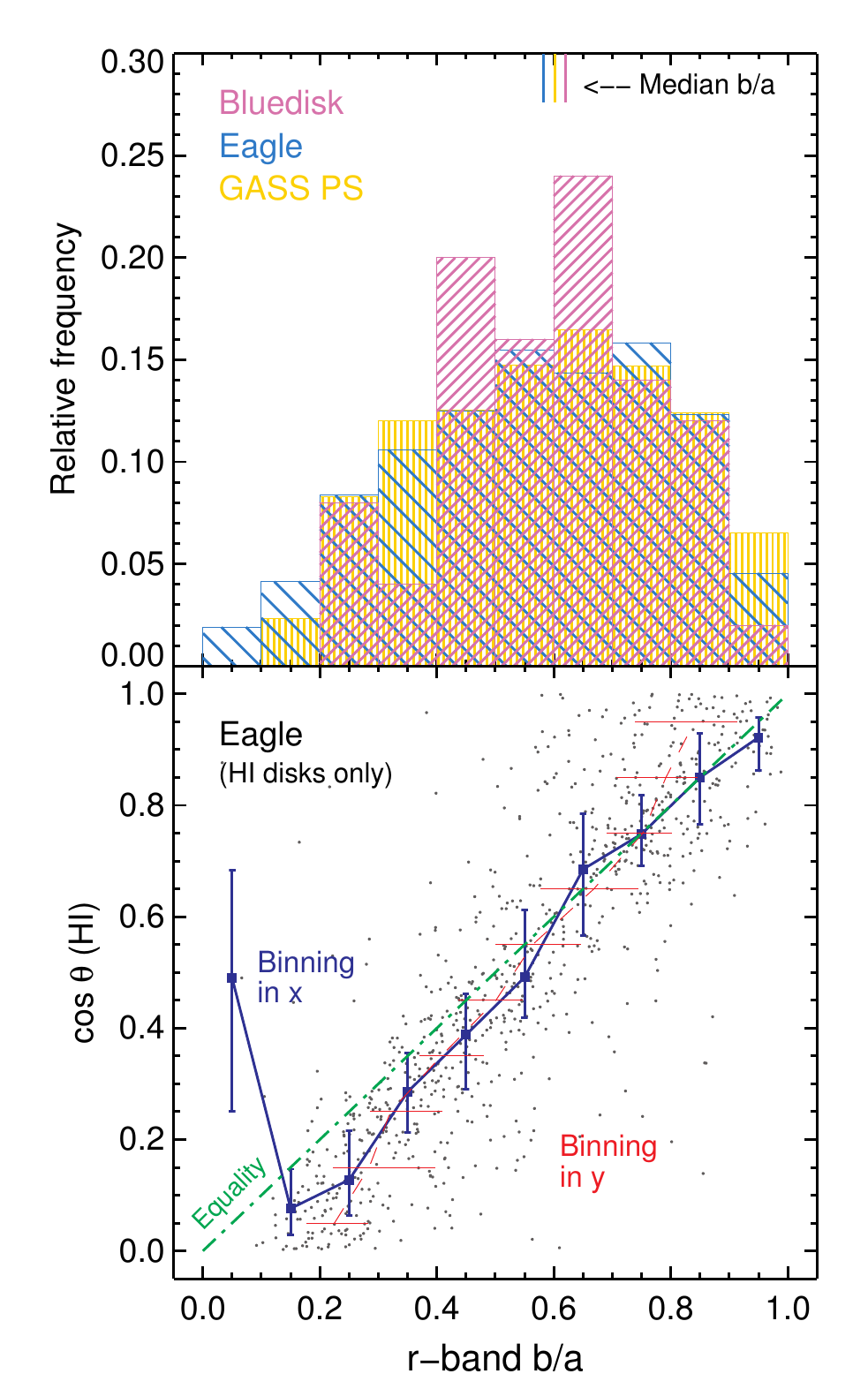}
       \caption{Correlation between the \hi{} disk inclination and stellar r-band ellipticity. The \textbf{top} panel shows the distribution of the latter quantity ($b/a$) for \bluedisk{} (magenta), the \gass{} parent sample (yellow), and \eagle{} (blue): This should be uniform if it actually measured the (random) inclination, but is clearly concentrated towards intermediate values of $\sim$0.6 in all samples (vertical ticks at the top indicate the medians, which agree well). \textbf{Bottom:} Direct comparison between $b/a$ and \hi{} inclination angle cos($\theta$) for \eagle{} galaxies with an \hi{} disk. There is a fair amount of scatter for individual galaxies, but for $b/a \gtrsim 0.6$ the approximation that $\cos(\theta) = b/a$ is true on average.}
    \label{fig:app.incli}
  \end{figure}

In the top panel of Fig.~\ref{fig:app.incli} we show the distribution of axis ratios $b/a$ fit to the 25 mag arcsec$^{-2}$ r-band isophote of our \eagle{} galaxies (blue) and those for the \bluedisk{} sample (magenta), as well as the \gass{} parent sample (yellow). Because the orientation of the galaxies in the simulation box is random, the distribution of inclination angles cos($\theta$) is flat, and the same should be true for $b/a$ if the two are equal in a statistical sense. However, this is clearly not the case for either simulations or observations: both show a marked deficit at both small ($b/a \lesssim 0.3$, i.e.~very elongated isophotes) and large ellipticities ($b/a \gtrsim 0.9$, almost circular). In principle, the \bluedisk{} sample selection could lead to a subtle selection bias towards certain axis ratios, but the fact that the distribution is very similar to the stellar-mass-only selected \eagle{} and \gass{} samples makes this very unlikely. A two-sided KS test shows a likelihood of 0.26 for the \eagle{} and \bluedisk{} distributions being drawn from the same parent population, and the medians -- vertical lines at the top of the plot -- are also very similar for all three data sets (0.58, 0.60, and 0.62 for \eagle{}, \gass{}, and \bluedisk{}, respectively). Instead, the uneven distribution simply reflects the fact that extreme axis ratios are rare because even if a (stellar) disk is aligned perfectly edge-on (face-on), any deviation from a circular, infinitely thin disk will increase (decrease) the measured axis ratio and therefore drive the $b/a$ distribution towards intermediate values.

In the bottom panel, we directly compare r-band axis ratio and \hi{} disk inclination angle for \eagle{} galaxies with an \hi{} disk (i.e.~excluding irregular \hi{} morphologies; grey points), and also show the running median and 25th/75th percentile binned both by $b/a$ (blue) and cos($\theta_\hi$) (red). There is clearly substantial scatter in the relation, of typically $\sim$0.1 around the median, but in general there is a clear correlation between the two. As a function of $\cos \theta$ (red), the relation is slightly steeper than one-to-one, which explains the uneven shape of the distributions in the top panel. More importantly for the interpretation of observational data, however, is the opposite trend: at $b/a \gtrsim 0.6$, the median $\cos\theta$ (blue) is almost exactly equal to $b/a$ (green line). Although care must evidently be taken when applying this relation to individual galaxies, the estimation of inclination angles from optical ellipticities should at least be reliable in a statistical sense for galaxies with moderately elliptical isophotes.


\section{Sensitivity of \hi{} profiles to the adopted beam size}
\label{sec:app.beamsize}
\begin{figure}
  \centering
  \includegraphics[width=\columnwidth]{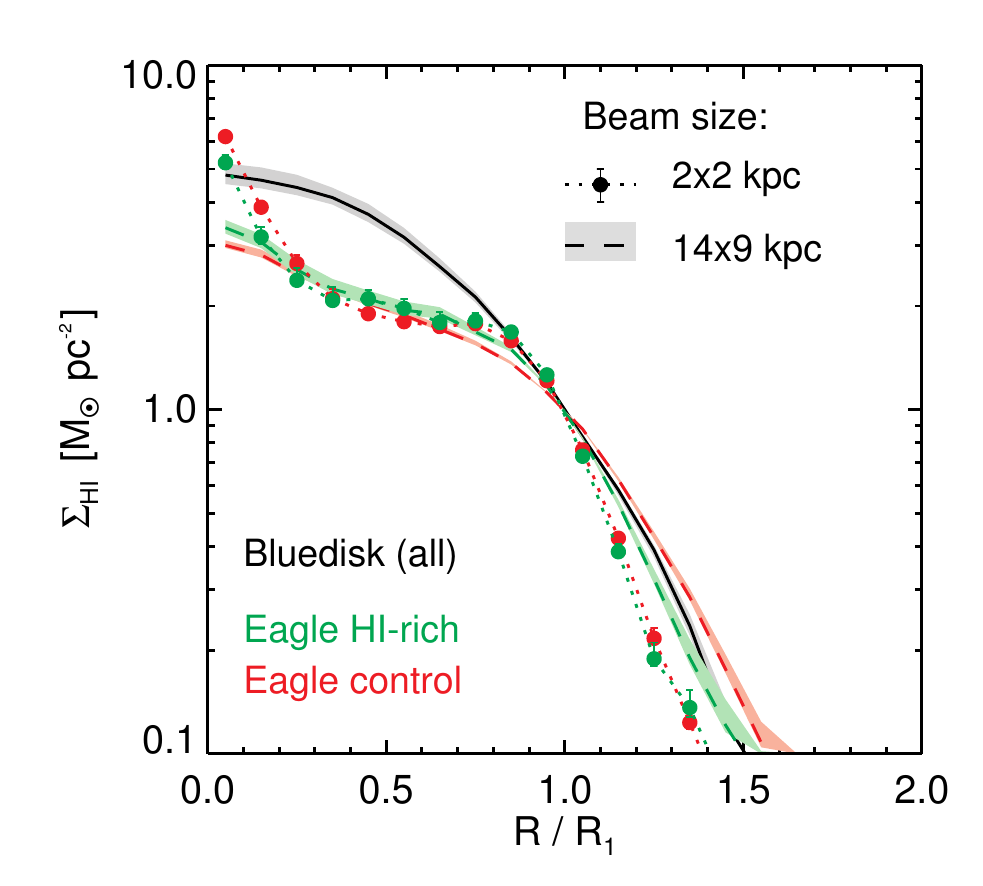}
  \caption{Variation of \hi{} surface density profiles with beam size. The default low-resolution profiles with beam size $14 \times 9$ kpc FWHM (see Section \ref{sec:profiles}) are shown with dashed lines, and the profiles obtained from higher-resolution maps with beam size $2\times 2$ kpc FWHM as filled circles connected with dotted lines. In both cases, red represents control galaxies and green \hi-rich galaxies. For guidance, the combined median \bluedisk{} profile is shown in black. Error bars indicate $1\sigma$ uncertainties on the median. Although there is some change in the detailed profile shape with increased resolution, the close match between \hi-rich and control galaxies is robust to this change.}
   \label{fig:bd.profiles_hires}
\end{figure}

In the profile comparison in Section \ref{sec:profiles}, we had artificially reduced the resolution of the simulated \hi{} maps to match the beam size of the \emph{Westerbork Synthesis Radio Telescope} (WSRT). We now briefly investigate how this smoothing has affected our results, by analysing the profiles obtained from higher-resolution images with a (circular) beam of FWHM = 2 kpc, more similar to what is shown in Fig.~\ref{fig:bd.visual_images}. Based on our discussion in Section \ref{sec:profiles}, we choose to rotate galaxies to face-on as the most physically meaningful way of extracting surface density profiles, and present the result in Fig.~\ref{fig:bd.profiles_hires}. Particle \hi{} masses are here calculated with the empirical BR06 H$_2$ formula.

The higher-resolution profiles are shown with filled circles connected by dotted lines, red for control galaxies and green for those which are \hi-rich. For comparison, we also re-produce the low-resolution ($14 \times 9$ kpc) profiles shown in Fig.~\ref{fig:bd.profiles} with dashed lines. Statistical $1\sigma$ uncertainties on the medians are indicated with error bars in the former case, and the width of the shaded bands in the latter. For guidance, we also include the overall \bluedisk{} median profile (black).

Unsurprisingly, the increase in resolution does lead to some change in the detailed shape of the profile, but overall the effect is rather small. In the outer regions ($R \gtrsim 0.9 R_1$), the high-resolution profile is slightly steeper (reaching $0.1 \msun\, \text{pc}^{-2}$ at 1.4 instead of $1.5\, R_1$); the same is true for the very centre ($R \lesssim 0.3\, R_1$). More interestingly, however, Fig.~\ref{fig:bd.profiles_hires} clearly demonstrates that the close agreement between \hi-rich and control galaxies discovered by \citet{Wang_et_al_2014} is \emph{not} sensitive to the beam size, and is a genuine physical feature of our simulations. This strongly suggests that the same should also be true for the real Universe, a prediction that can be tested directly in future with high-resolution \hi{} surveys such as from the \emph{Square Kilometer Array} (SKA).

\end{appendix}

\end{document}